\documentclass[12pt,a4paper]{article} 
\usepackage{color}
\usepackage{amsmath}
\usepackage{gensymb}
\usepackage{upgreek}
\usepackage{graphicx}
\usepackage{float}
\usepackage[margin=0.5in]{geometry}

\begin{document}
\title{Corner Reflector Plasmonic Nanoantennas for Enhanced Single-Photon Emission}
\author{P. Chamorro-Posada\\
  Dpto. de Teor\'{\i}a de la Se\~nal y Comunicaciones\\
 e Ingenier\'{\i}a Telem\'atica,\\
 Universidad de Valladolid, ETSI Telecomunicaci\'on,\\
 Paseo Bel\'en 15, 47011 Valladolid, Spain}
\maketitle

\begin{abstract}
  The emission rate of atom-like photon sources can be significantly improved by coupling them to plasmonic resonant nanostructures. These arrangements function as nanoantennas, serving the dual purpose of enhancing light--matter interactions and decoupling the emitted photons. However, there is a contradiction between the requirements for these two tasks. A small resonator volume is necessary for maximizing interaction efficiency, while a large antenna mode volume is essential to achieve high emission directivity. In this work, we analyze a hybrid structure composed of a noble metal plasmonic resonant nanoparticle coupled to the atom-like emitter, which is designed to enhance the emission rate, alongside a corner reflector aimed at optimizing the angular distribution of the emitted photons.  A comprehensive numerical study of silver and gold corner reflector nanoantennas, employing the finite difference time domain method, is presented.  The results demonstrate that a well-designed corner reflector can significantly enhance photon emission directivity while also substantially boosting the emission rate.
  \end{abstract}

\section{Introduction}

Efficient and reliable single-photon sources, capable of producing on-demand high-purity single-photon states, are key elements within the emerging field of photon-based quantum technologies \cite{flamini,divincenzo}.  Atom-like emitters, based either on solid-state or molecular systems, are a promising alternative for this purpose \cite{aharonovich2016,bogdanov,toninelli2021,dibos2018,gaither}.  In the past years, nanoscale elements have been introduced to enhance the emission rates of photon sources and to tailor their spatial radiation properties.  Such nanoantennas (NAs) reproduce at the nanoscale the action of traditional radio frequency (RF) aerials. Various traditional antenna geometries have been considered for applications in the optical realm: dipole \cite{dipole,chowdhury}, \mbox{patch \cite{patch}}, \mbox{bowtie \cite{bowtie}} or the Yagui--Uda array \cite{novotny2011,yagui1,yagui2}.

Antennas for light \cite{novotny2011} serve the dual purpose of enhancing emitter fluorescence and facilitating the outcoupling of the generated photons.  The performance requisites for the two functions are antithetical:  whereas a small mode volume is required for an enhanced emission rate, high antenna directivity is well known to be the result of a large electrical size, i.e, the extent of the structure measured using the wavelength of the electromagnetic field as unit reference \cite{krauslibro,balanis}.  In  \cite{bogdanov}, such antipodal requirements have been reconciled by photomodifying a nanopatch antenna assembled around a nanodiamond with a nitrogen-vacancy center to reduce the cavity mode size while keeping a large size emitter to spatially shape the radiation produced. In this work, we tackle this issue by using a hybrid plasmonic structure built from the combination of a conventional NA acting as the feeder element of a corner reflector (CR) \cite{kraus}.  This strategy simultaneously permits optimization of the emission rate at the feeder design and the capacity to use a CR to enhance the directivity of the \mbox{photons produced}.

The use of CRs in antenna design has been prevalent since Kraus introduced them in 1940 \cite{kraus}, owing to their outstanding performance, simplicity, and ease of construction.  This work demonstrates that these advantageous properties can also be applied to nanoscale photonics. The potential of CRs to enhance photon emission was highlighted in \cite{chamorro}, where the radiation patterns generated by an electric dipole emitter with a perfect conductor CR were analyzed.  Here, we present a thorough numerical study of the proposed NA, utilizing two alternative noble metals: silver and gold.  The feeder is assumed to be constituted by an atom-like dipole emitter coupled to a spherical nanoparticle (NP), though the proposed CRs can also accommodate more complex feeder geometries, such as dimers or bowtie structures \cite{dipole,bowtie}, which may offer enhanced emission.  The analysis is conducted using the finite difference time domain method (FDTD) \cite{yee, taflove}, enabling the calculation of the radiative emission rate enhancement factor, losses within the metallic structures, and the far-field patterns of the emitted photons. This approach provides a comprehensive characterization of the NA.   

\begin{figure}[H]
  \includegraphics[width=11.5 cm]{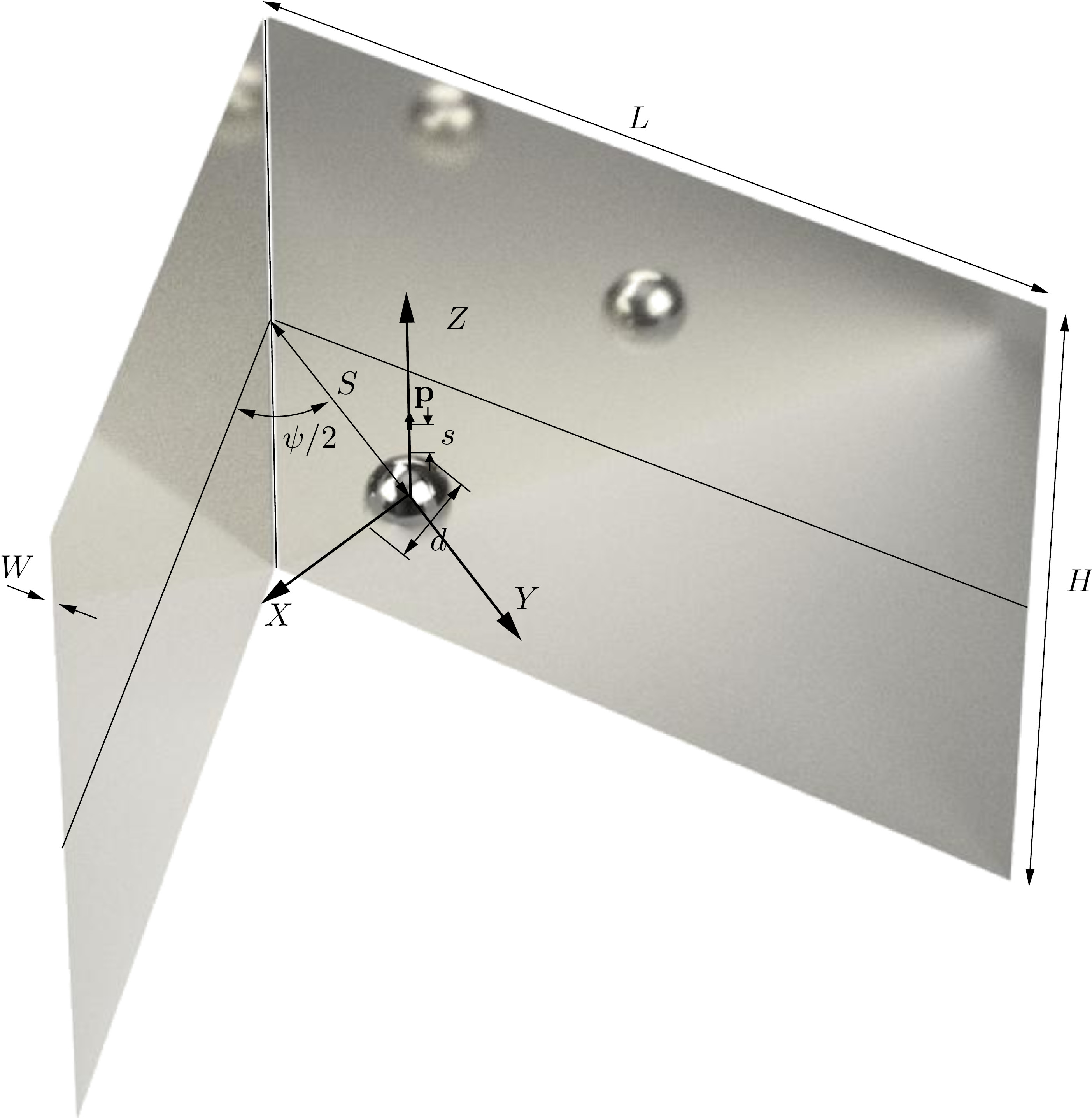}
\caption{Geometry employed  in the analyses.  The feeder is composed by a dipole emitter {$\mathbf{p}$} coupled to a nanosphere of diameter $d$.  The corner, of angle $\psi$, is assumed to be built from to sheets of the same metal as the NP with dimensions $L\times H$.  \label{fig::geometry}}
\end{figure}

 The geometry employed in the analyses is depicted in Figure \ref{fig::geometry}.  A spherical NP with diameter $d$ is positioned at the origin.  The atom-like emitter is represented by a point dipole source $\mathbf{p}$  located at a distance $s$ from the surface of the metal nanosphere that is placed at a distance $S$ from the corner apex along the bisector.  The corner angle is $\psi$, while the metal films have a length $L$ and height $H$, as shown in the figure. The thickness of the metal films forming the reflector is $W$. The planes of the CR are assumed to be made of the same metal as used for the NP in the feeder system.

A metallic plane typically reflects a substantial portion of an incident wave originating from a large distance, due to the significant mismatch between its optical properties and those of the surrounding dielectric.  However, when a noble metal plane is placed near a dipolar source, it can significantly influence the emission rate \cite{novotny2006} in a similar manner to the behavior observed with emission near NPs. Therefore, analyzing only the radiation pattern of emitted photons is insufficient for designing CR-based NAs. A comprehensive analysis, including the effects on radiative emission and loss rates, has been conducted. The impact of the thickness of the metal films on these properties has also been addressed.

\section{{Materials and Methods}}\label{MM} 

The numerical calculations were conducted using ANSYS Lumerical FDTD Solutions,  release 2024 R2.1, on a HP Z6 G4 Workstation with two Intel Xeon 6226R processors and 512 GB of RAM. For the FDTD simulations, a variable-size discretization was employed, with a finer grid in the dipole-nanoparticle region. The discretization parameters were adjusted based on a series of convergence analyses, setting the maximum grid step to $\Delta l=1$ nm within this region.  To reduce computational costs, the symmetry planes $XZ$ and $YZ$ were utilized for the feeder system, and $YZ$ symmetry was exploited in simulations, including the CR. In principle, exploiting the symmetry of the problem when specifying the boundary conditions should not affect the accuracy of the obtained solutions, which should remain consistent with the symmetry considerations, regardless of the boundary conditions used. Nevertheless, several test computations were carried out without the symmetry boundary conditions, and no significant differences were observed in the results.  The dipole source emitted a short pulse covering the spectral region between $300$ and $800$ nm. The total duration of the simulations was set to $T=150$ fs.  The analyses focused on a target wavelength $\lambda_{Ag}=450$ nm for silver and $\lambda_{Au}=600$ nm for gold. The wavelength of interest in any application will depend on the specific fluorophore being used. To ensure a fair comparison between the two metals, which have different plasmonic resonance wavelengths, the target wavelengths for the silver and gold NAs were arbitrarily selected to be redshifted from their respective resonances by similar spectral separations, typical of those found in applications.

In spite of the intrinsically quantum nature of the fluorescence effect,  the enhancement factor of the radiative decay rate, $f_R$, due to the Purcell effect \cite{purcell} of a nanoantenna in an homogeneous dielectric environment can be calculated using a fully classical \mbox{model \cite{novotny2006,krasnok}} as the ratio between the total power collected in the far-field region when the noble metal antenna is present, $P_{rad}$, and the power that would be radiated from a point dipole source in free space, $P_0$:
\begin{equation}
  f_R=\dfrac{\gamma_r}{\gamma_{r,0}}=\dfrac{P_{rad}}{P_0},\label{eq::fr}
\end{equation}
where $\gamma_{r}$ ($\gamma_{r,0}$) is the radiative emission rate in the presence (absence) of the plasmonic structure.  $P_{rad}$ and the radiative emission rate enhancement factor, or Purcell factor, are readily obtained from FDTD computations \cite{kaminski,chowdhury}.  For this purpose, the atom-like emitter was modeled using a time-windowed dipole soft source.  Instead of setting a given value of the electromagnetic field at a point in space, the field from this type of source adds to the existing electromagnetic fields without causing scattering, therefore permitting the modeling of fluorescence.   The value of $P_0$ can be computed from its analytical \mbox{expression \cite{jackson}}, but it is also provided by Lumerical via the  sourcepower  script command.  On the other hand, complex non-exponential or non-Markovian emission processes cannot be accounted for by the model. 

Similarly, the impact of the losses in the antenna structure, which affect the overall quantum efficiency, can be evaluated from the normalized loss rate factor
\begin{equation}
  f_L=\dfrac{\gamma_{l}}{\gamma_{r,0}}=\dfrac{P_{loss}}{P_0}.\label{eq::fl}
\end{equation}
$P_{loss}$  is computed as $P_{loss}=P-P_{rad}$, where $P$ is the power radiated by the dipole in the NA environment used to modify its radiation in a vacuum---$P_0$.

The total photon emission rate is influenced by radiative and non-radiative transitions and losses in the noble metal structures.  The emission yield is characterized by the quantum efficiency $\eta_Q$ of the process: the ratio between the rate of radiative transitions and the total transition rate, including radiative $\gamma_r$ and non-radiative $\gamma_{nr}$ transitions, and the rate of photons lost in the metals $\gamma_l$ \cite{novotny2006}, 
\begin{equation}
  \eta_Q=\dfrac{\gamma_{r}}{\gamma_{r}+\gamma_{nr}+\gamma_l}.\label{eq::qe}
  \end{equation}

If we assume a highly efficient atom-like emitter, with negligible non-radiative transitions $\gamma_{nr}<<\gamma_{r}$, the quantum efficiency can be approximated as
\begin{equation}
    \eta_Q\simeq\dfrac{\gamma_{r}}{\gamma_{r}+\gamma_l}=\dfrac{f_R}{f_R+f_L}, \label{eq::qe2}
\end{equation}
where we have divided both numerator and denominator of the above expression by $\gamma_{r,0}$.  If we substitute \eqref{eq::fr} and \eqref{eq::fl} into \eqref{eq::qe2}, the resulting expression coincides with the classical definition of the radiation efficiency of an antenna \cite{krauslibro,balanis}:
\begin{equation}
  \eta=\dfrac{P_{rad}}{P_{rad}+P_{loss}}.
  \end{equation}

The evaluation of the emission properties for the NAs in this study is based on the calculation of $f_R$ and $f_L$ from the FDTD results to gauge the photon emission rate.  Both gold and silver plasmonic NA were considered, with their respective optical constants obtained from \cite{palik,johnson}.

The external photon collection capabilities of the hybrid system were evaluated using on-axis directivity, which is determined from the far-field radiation pattern as
\begin{equation}
  D=4\pi\dfrac{K_0}{P_{rad}},\label{eq::D}
\end{equation}
where $K_0=K\left(\theta=\pi/2,\phi=\pi/2\right)$ is the radiation intensity emitted along the $Y$ axis, i.e., the power-per-unit solid angle in that direction.  In Equation \eqref{eq::D}, $D$ measures the ratio of the power radiated per unit solid angle along the $Y$ axis to that of a  hypothetical isotropic radiator \cite{krauslibro,balanis}.

The antenna gain, defined as
\begin{equation}
  G=\eta D, \label{eq::gain}
  \end{equation}
combines the emission efficiency $\eta$ and directivity $D$, serving as the key metric for antenna performance in an RF link \cite{krauslibro,balanis}.  This figure of merit is applicable to optical NAs, where it indicates the potential for photon collection from single-photon emitters.  We now analyze the values of $D$, $f_R$, and $f_L$, which are the three parameters that determine $G$ for silver and gold NAs.     

The Kildal limit \cite{kildal} sets a theoretical upper bound on the maximum directivity that can be achieved with an antenna
\begin{equation}
  D_{max}=(k R)^2+3, \label{eq::kildal}
  \end{equation}
where $k$ is the wavenumber, and $R$ is the radius of the sphere circumscribing the antenna. Antennas having directivities beyond the limits imposed by their physical dimensions are often named superdirective antennas \cite{hansen}.  It was early realized that the implementation of such antennas is subject to physical restrictions \cite{chu,harrington} such us very high values of $Q$, i.e.,  very low bandwidths, that are also associated with very low efficiencies due to very large ohmic losses, as well as very sensitive implementation tolerances \cite{hansen}.
 
Antenna designs based on dielectric resonators coupled to a dipole emitter have proven particularly advantageous in the pursuit of electrically small, practical superdirective antennas with high efficiency. This design has been theoretically analyzed at the nanoscale \cite{krasnok2} and experimentally demonstrated in the UHF band \cite{gaponenko}. It is also well established that superdirectivity is not viable for small plasmonic nanoparticles \cite{krasnok2}; however, in principle, plasmonic nanostructures can outperform their dielectric counterparts despite the intrinsic metal losses they exhibit \cite{khurgin}. The approach in this work aligns with classical antenna design, where using a reflector provides a straightforward solution to spatially shaping the emitted field into a highly directive beam by enlarging the antenna \mbox{dimension $R$.}

\section{Results and Discussion}

An initial analysis was performed on the isolated dipole-NP feeder problem.  This permitted the full evaluation of the impact on the emitter properties of the incorporation of the CR.  To this end, a series of FDTD calculations were performed on the geometry displayed in Figure \ref{fig::geometry} in the absence of the metal sheets defining the CR.

\begin{figure}[H]   
    \begin{tabular}{cc}
    {(a)}&{(b)}\\
    \includegraphics[width=6.5 cm]{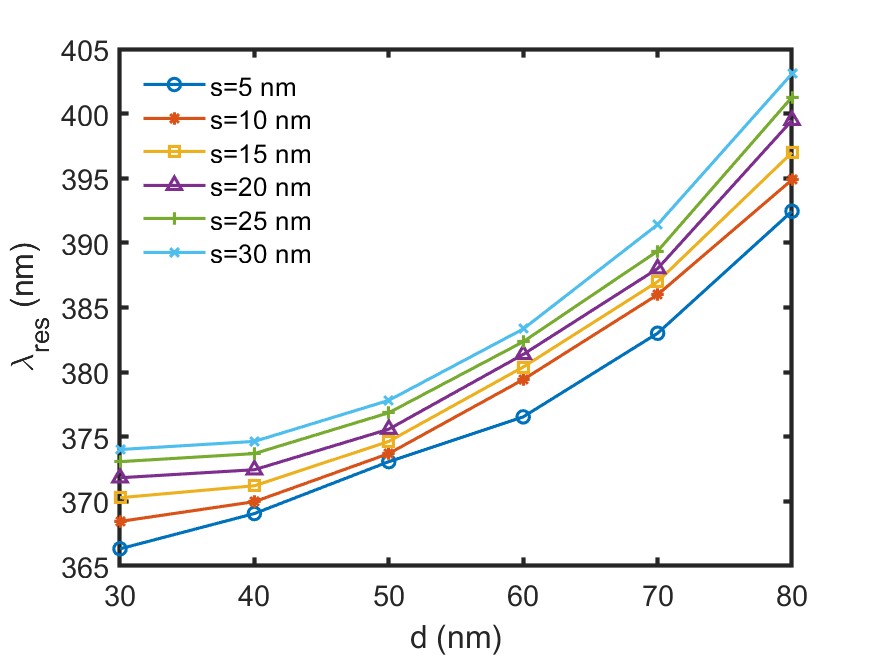}&
    \includegraphics[width=6.5 cm]{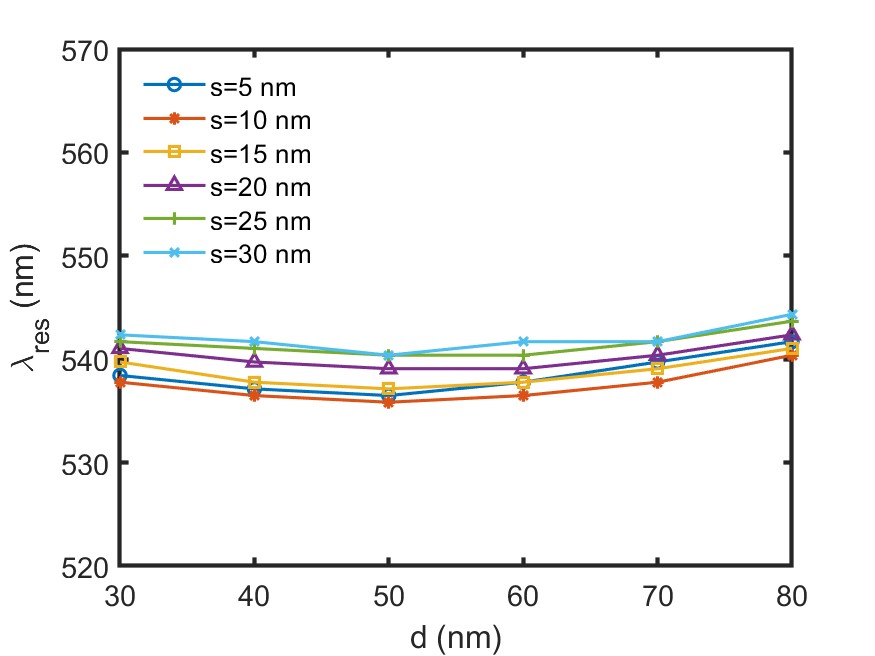}
    \end{tabular}
\caption{Calculated resonance wavelengths of silver (\textbf{a}) and gold (\textbf{b}) NPs as a function of their diameters. \label{fig::resonancias}}
\end{figure}

The resonances in the spectral responses of NPs are responsible for their characteristic colors, and their intensities and frequencies depend on the particle size, shape, and the dielectric properties of the surrounding medium.  Figure \ref{fig::resonancias} displays the resonances calculated from the emission spectrum of the corresponding NPs coupled to the dipole emitter.  For gold NPs surrounded by air, the calculations show a negligible displacement of the plasmon resonance for the range of sizes considered, which is in accordance with previous calculations \cite{liz}.  Furthermore, the spectral shift of the resonance with the particle size correlates well with the redshift of the resonance, as increased particle size was observed in measurements performed for colloidal suspensions of silver NPs of sizes varying between $60$ and $80$ nm in \cite{alqadi,qin}.  To obtain the data in Figure \ref{fig::resonancias}, the smooth spectra from FDTD simulations were Nyquist interpolated to increase the spectral resolution without altering the information contained in the original data.

\begin{figure}[H]  
  \begin{tabular}{cc}
    {(a)}&{(b)}\\
    \includegraphics[width=6 cm]{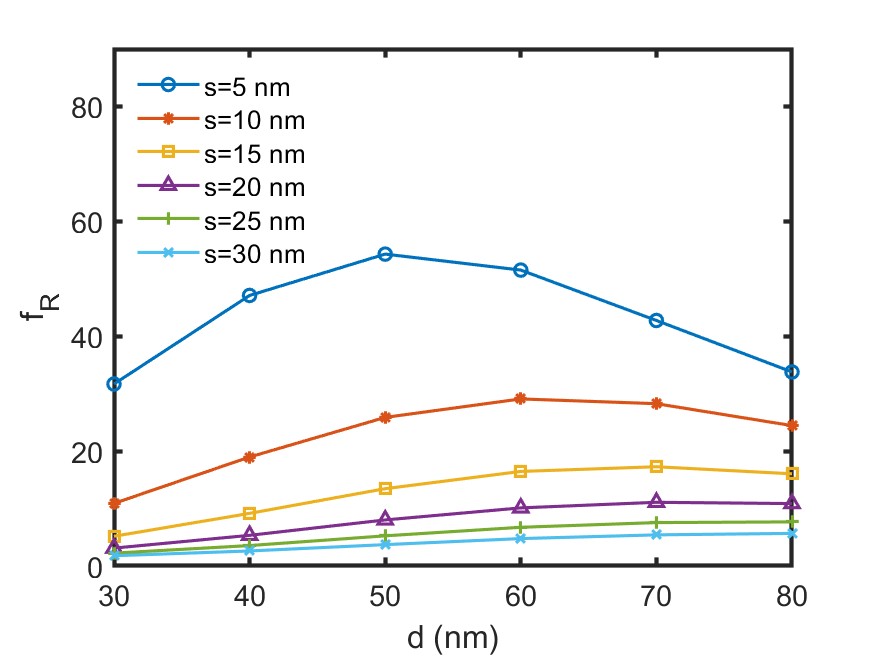}&
    \includegraphics[width=6 cm]{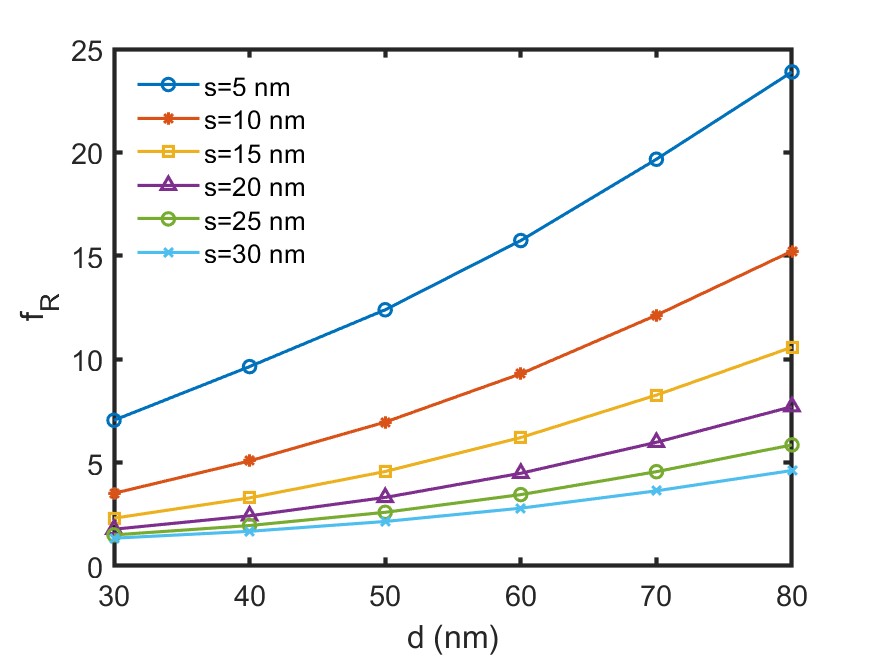}\\
    {(c)}&{(d)}\\
    \includegraphics[width=6 cm]{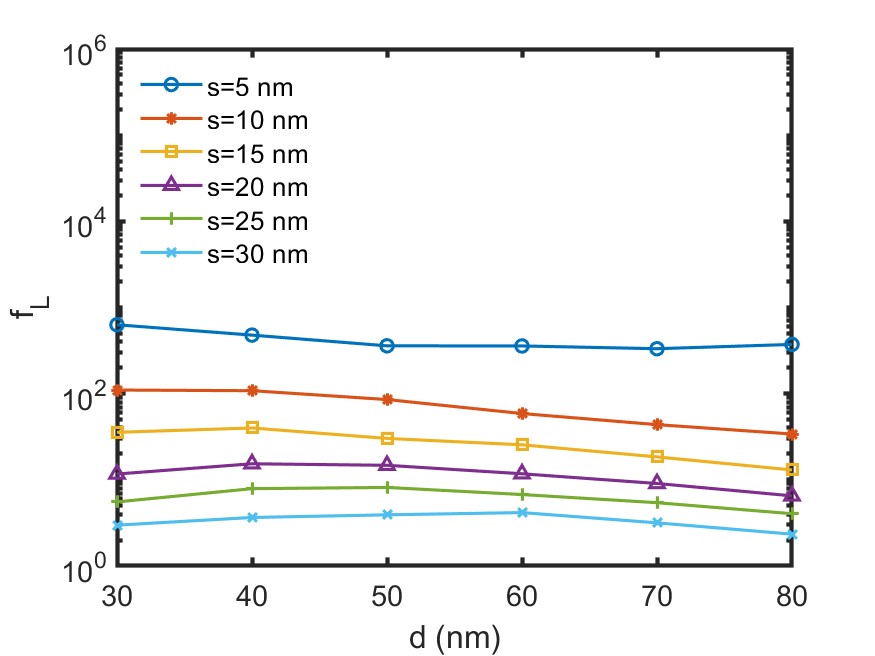}&
    \includegraphics[width=6 cm]{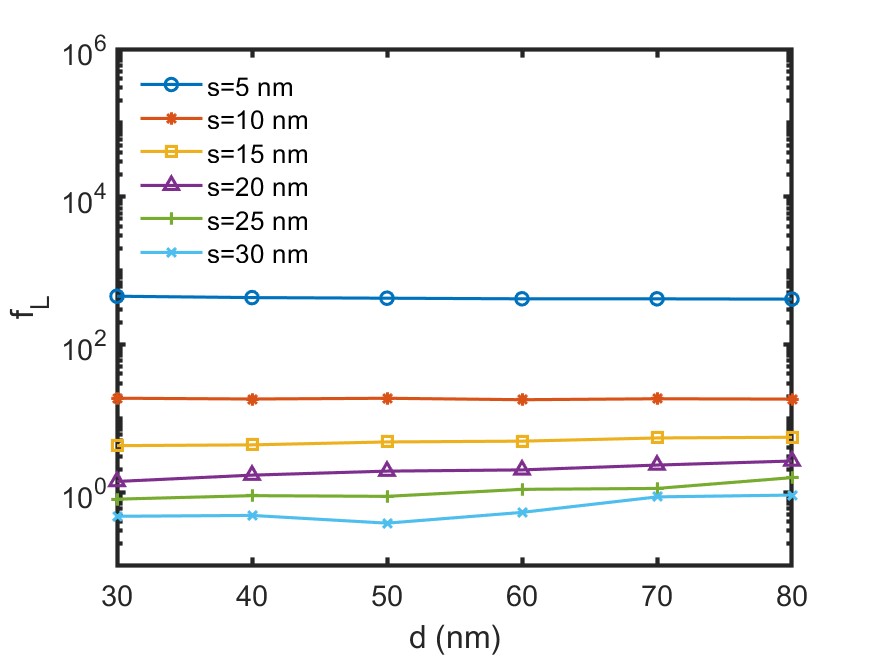}
    \end{tabular}
\caption{Purcell, $f_R$, and loss, $f_L$, factors of silver NPs calculated for different spherical particle diameter values $d$ and separation values $s$ from the the dipole emitter.  (\textbf{a},\textbf{c}) plots are the values at resonance and, (\textbf{b},\textbf{d}) are at the target wavelength of $\lambda_{Ag}=450$ nm.\label{fig::silverNPs}}
\end{figure}

\begin{figure}[H]    
    \begin{tabular}{cc}
    {(a)}&{(b)}\\
    \includegraphics[width=6 cm]{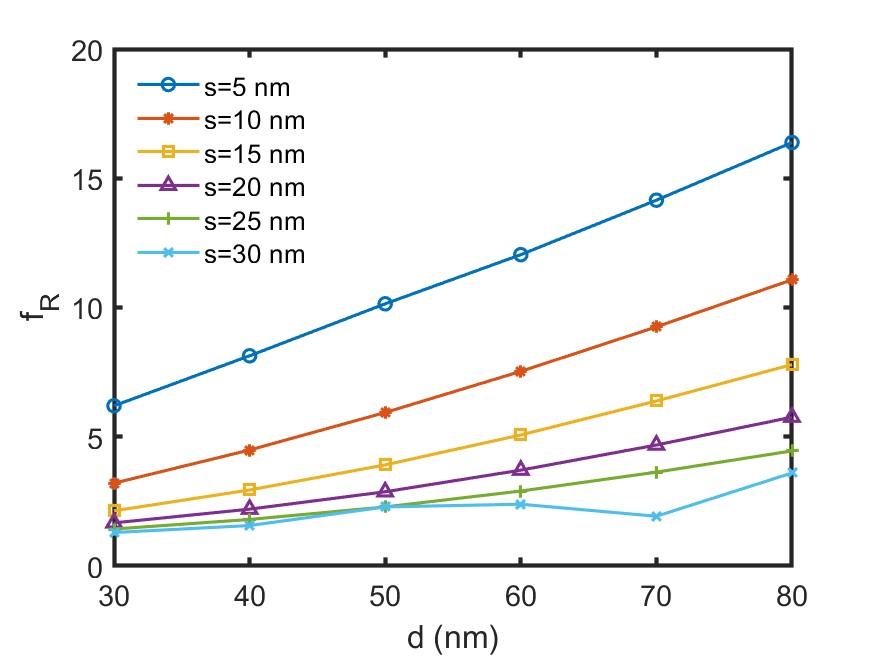}&
    \includegraphics[width=6 cm]{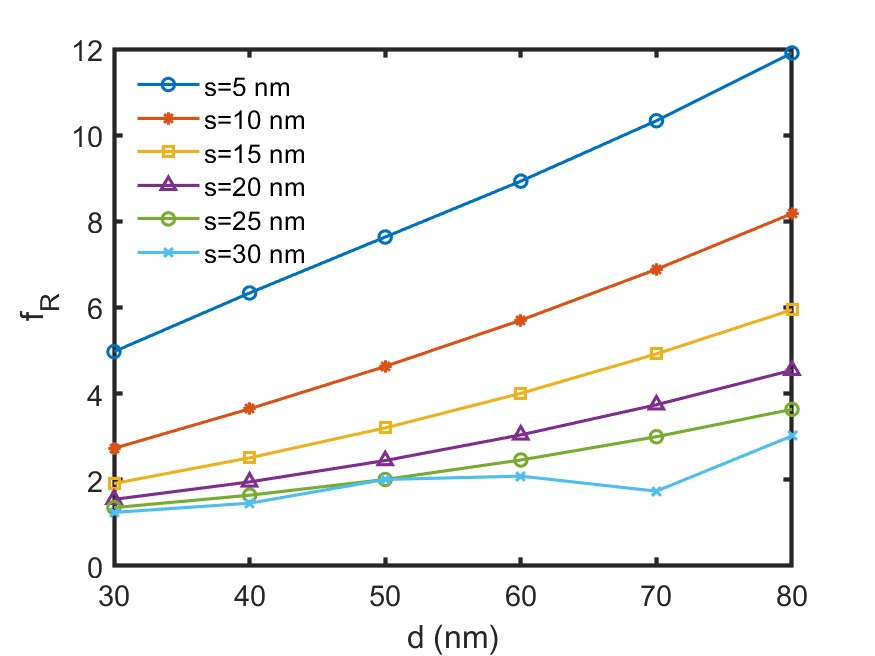}\\
    {(c)}&{(d)}\\
    \includegraphics[width=6 cm]{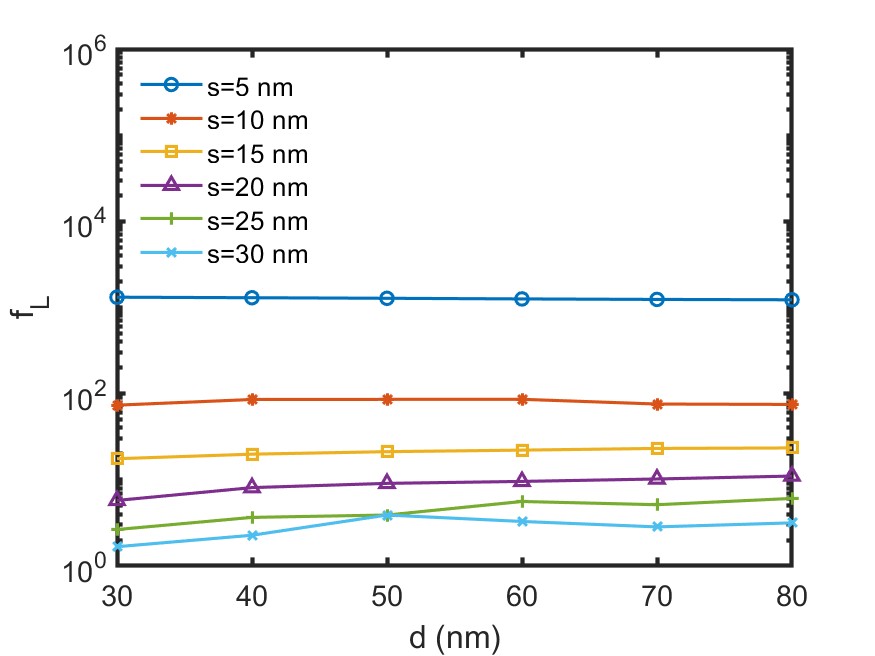}&
    \includegraphics[width=6 cm]{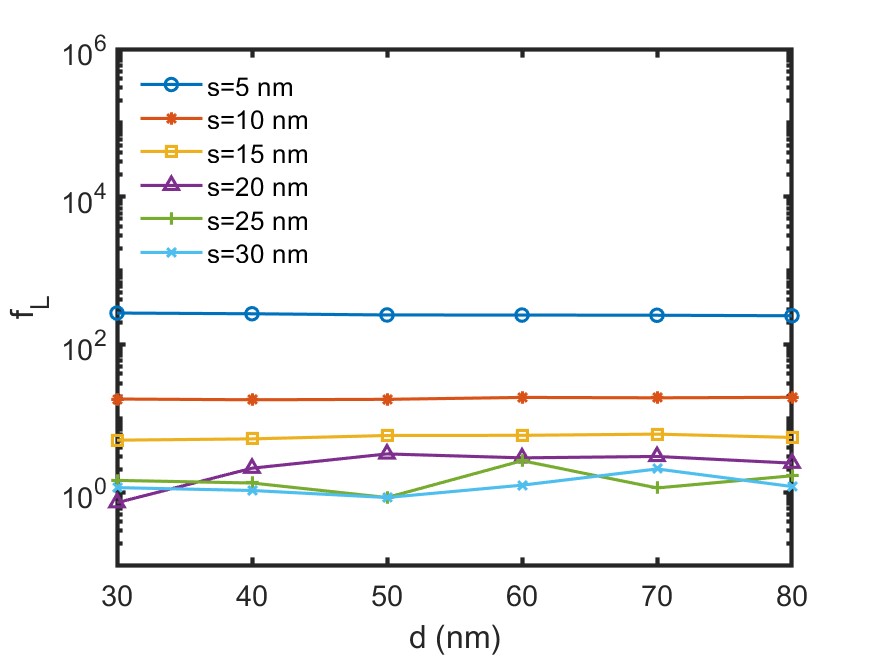}
    \end{tabular}
\caption{Purcell, $f_R$, and loss, $f_L$, factors of gold NPs calculated for different spherical particle diameter values $d$ and separation values $s$ from the the dipole emitter.  (\textbf{a},\textbf{c}) plots are the values at resonance, and (\textbf{b},\textbf{d}) are at the target wavelength of $\lambda_{Au}=600$ nm.\label{fig::goldNPs}}
\end{figure}

Figure \ref{fig::silverNPs} displays the results obtained for the normalized factors $f_R$ and $f_L$ obtained from FDTD calculations for silver NPs of diameter $d$ coupled to a dipole emitter at a distance $s$ and perpendicular to a spherical NP.  Particle sizes between $d=30$ and $80$ nm were considered, as well as dipole separations ranging from $s=5$ nm to $s=30$ nm.  The plots display data collected at the resonance and at the target wavelength of  \mbox{$\lambda_{Ag}=450$ nm}.  The corresponding results for gold NPs with their target wavelength at $\lambda_{Au}=600$ nm are shown in Figure \ref{fig::goldNPs}.  In this case, the calculations were performed for the same range of parameters $d$ and $s$.

\unskip

The calculated values of the radiative transition factor, $f_R$, at the target wavelengths generally increased as the particle size grew and the distance between the emitter and the NP diminished, both for silver and gold NPs.  However, the variation in terms of $f_R$ with particle size $d$ at resonance differed between silver and gold.  For gold NPs, the trend for $f_R$ at resonance was similar to that observed at $\lambda_{Au}$, while for silver NPs, $f_R$ exhibited a local maximum at a specific $d$ when the separation $s$ was small.  In general, $f_R$ reached higher values for silver NPs than for gold, both at resonance and at the target wavelength.  Although the relative size of the NPs compared to the wavelength at which $f_R$ was measured was larger for silver NPs, the ratio of the relative sizes was smaller than the observed ratios of $f_R$ values.  On the other hand, when the value of the wavelength was taken in consideration, the corresponding relative distances between the source and the NP were larger for silver NPs.

Regarding photon loss due to absorption at the NP, the value of $f_L$ increased significantly as the distance between the dipole emitter and the NP decreased.  Conversely, the NP size $d$ showed a lesser effect on $f_L$.  Overall, the loss factors for silver NPs were smaller, though still comparable, to those for gold NPs.

As discussed in Section \ref{MM}, achieving efficient emission requires a good balance between relatively low $f_L$ values and sufficiently high $f_R$ values.  Based on the analysis of the coupled dipole-nanoparticle system, a particle size of $d=70$ nm with a separation of $s=15$ nm from the dipole was chosen for the analysis of the full structure incorporating the CR. For these parameters, at their respective target wavelengths, the results show $f_L=5.4$ and $f_R=8.3$ for silver nanoparticles and $f_L=6.1$ and $f_R=4.9$ for gold nanoparticles.  This indicates a higher radiative emission rate enhancement with lower losses for silver compared to gold.

Under these conditions, $f_R>f_L$ for silver NPs, whereas $f_R<f_L$ for gold NPs.  The corresponding values of the emission gain figure of merit, $G$, as defined in \eqref{eq::gain}---with a directivity value of $D=1.5$ for an electrically short dipole \cite{krauslibro,balanis}---were $G_{Ag}=0.9$ for silver and $G_{Au}=0.7$ for gold NPs.

The radiation properties of the CRs in RF antennas were systematically studied in \cite{cottony} through extensive experimental measurements of the gain achieved with different corner configurations using a resonant half-wave dipole feeder.  As the distance from the corner apex to the feeder increased, a series of positions corresponding to local maxima of the antenna gain was identified. The first three positions were thoroughly analyzed in \cite{cottony}.

Whereas the second and third positions were sharply defined, the first position corresponded to a broader region.  This is due to a trade-off between the increase in directivity obtained by moving the feeder closer to the apex and the associated rise in ohmic losses in the conductors. In the first position, the corner angle for maximum gain depended on the width and length of the reflecting planes: as the planes became larger, the optimal angle decreased and the maximum gain increased.  Conversely, fixed optimal values of the corner angle were observed for the second and third positions when the planes were sufficiently large.  For the second position, a maximum gain occurred at $\psi=65^{\circ}$ when the width and length of the corner planes exceeded about one wavelength. At the third position, an optimal corner angle was found at $\psi=90^{\circ}$ when the reflecting planes were larger than half a wavelength.

In the following analyses, we selected a corner angle of $\psi=65^{\circ}$ and corner plane dimensions of $L=1.5$ $\upmu$m and $H=1$ $\upmu$m.  We examined the directivity $D$ along the corner axis and investigated how the presence of the CR influenced the values of $f_R$ and $f_L$ at the target wavelengths for silver and gold NAs.

\begin{figure}[H]   
    \begin{tabular}{cc}
    {(a)}&{(b)}\\
    \includegraphics[width=6 cm]{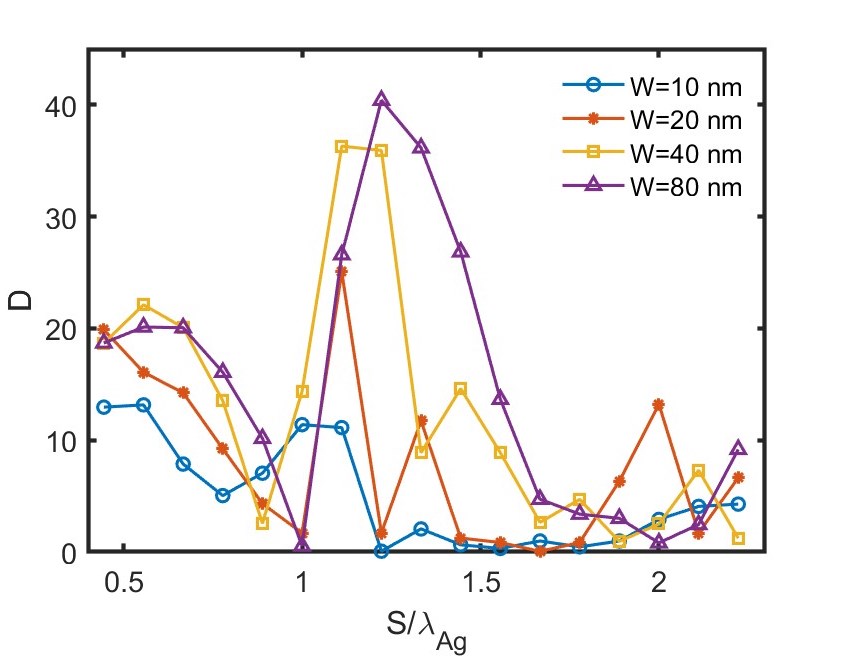}&
    \includegraphics[width=6 cm]{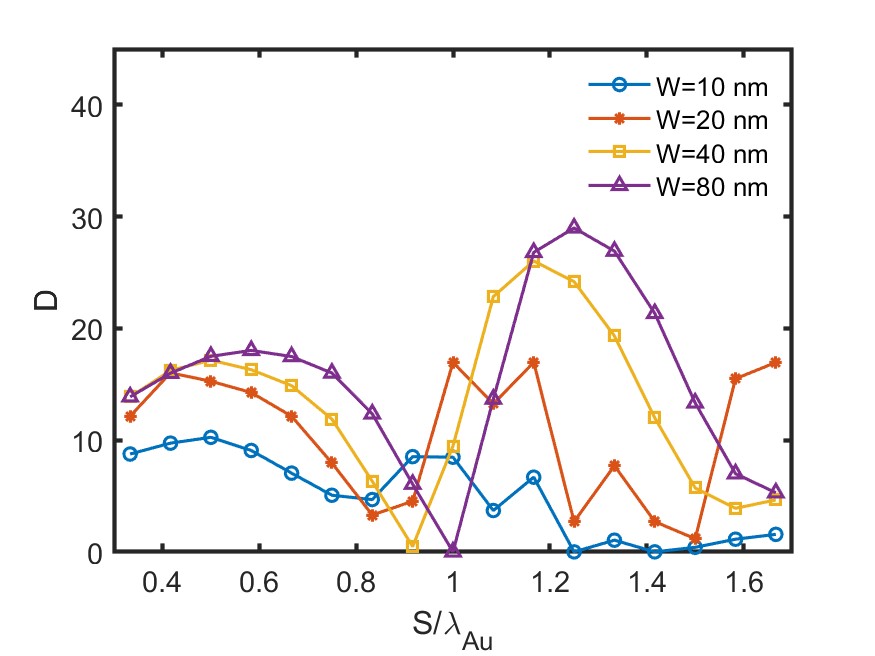}
    \end{tabular}
\caption{Values of the  directivity as a function of the feeder position $S$, normalized to the respective target wavelength, for four different reflector film thicknesses. (\textbf{a}) Results for the silver CR, with a target wavelength of $\lambda_{Ag}=450$ nm.  (\textbf{b}) Results for the gold CR, with a target wavelength of $\lambda_{Au}=600$ nm. \label{fig::directividades}}
\end{figure}

Figure \ref{fig::directividades} presents the on-axis directivities calculated for silver (a) and gold (b) CRs, with film thicknesses ranging from $W=10$ nm to $80$ nm and feeder positions between $S=0.2$ $\upmu$m and $S=1$ $\upmu$m.  For the presentation of the results, the values of $S$ have been rescaled with the respective target wavelengths for the gold and silver reflectors at which the directivities were evaluated.

The figure reveals similar patterns in the variation in directivity with the normalized separation $S/\lambda_M$ and the film thickness $W$, where $M$ denotes the corresponding noble metal.  The higher peak directivities obtained for silver than for gold can be attributed mainly to the larger relative values of the corner plane length $L/\lambda_{Ag}>L/\lambda_{Au}$.

The metal film thickness also played a significant role in the radiation properties of the NA.  As the thickness increased, the peak directivity increased, and the variation in  $S/\lambda_M$ followed a more regular oscillatory pattern in terms of the local maxima and minima, as in the case of RF antennas \cite{cottony}.  The irregularities in variations of directivity at smaller $W$ values were less pronounced for gold mirrors compared to silver.  Additionally, the positions of the maxima and minima shifted to larger values of  $S/\lambda_M$ as $W$ increased.  At the largest thickness of \mbox{$W=80$ nm}, the directivity was null precisely at $S=\lambda_M$. 

\begin{figure}[H] 
    \begin{tabular}{ccc} 
    \includegraphics[width=3.5 cm]{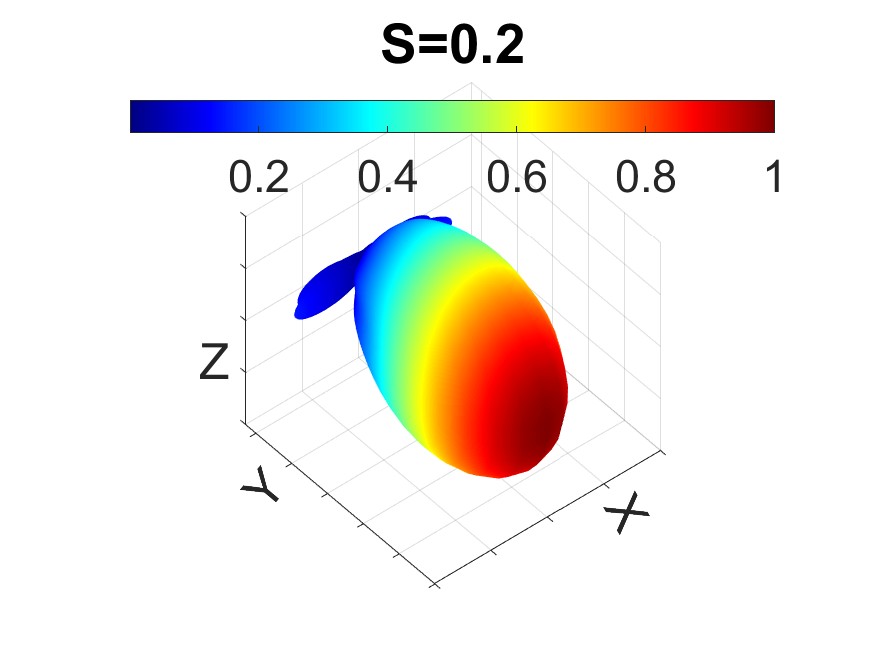}&
    \includegraphics[width=3.5 cm]{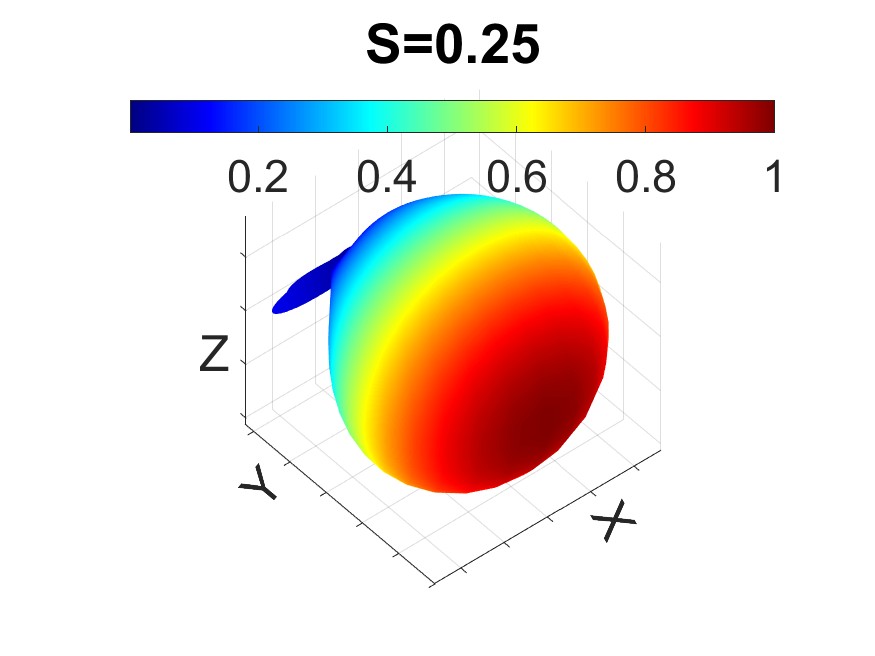}&
    \includegraphics[width=3.5 cm]{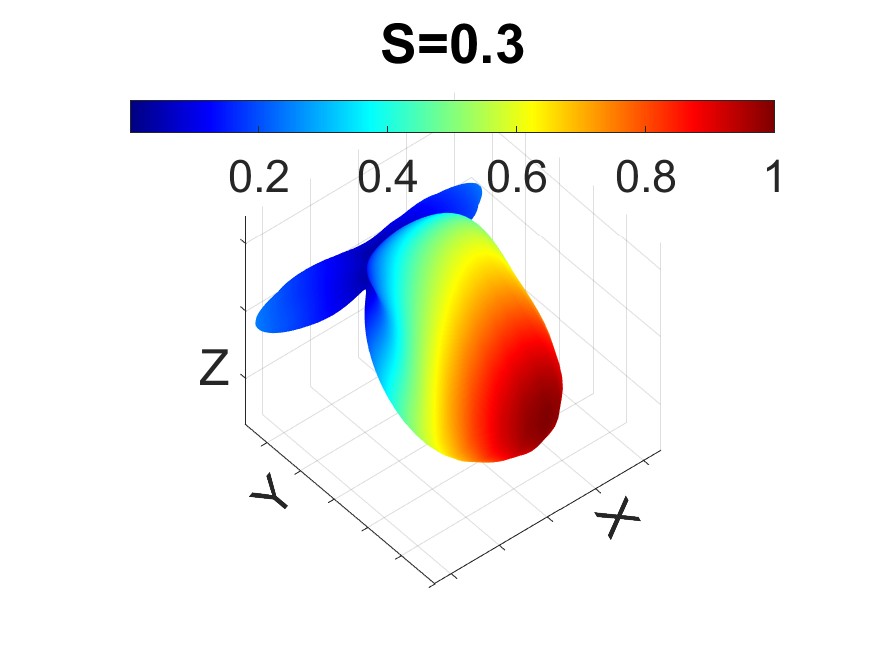}\\
    \includegraphics[width=3.5 cm]{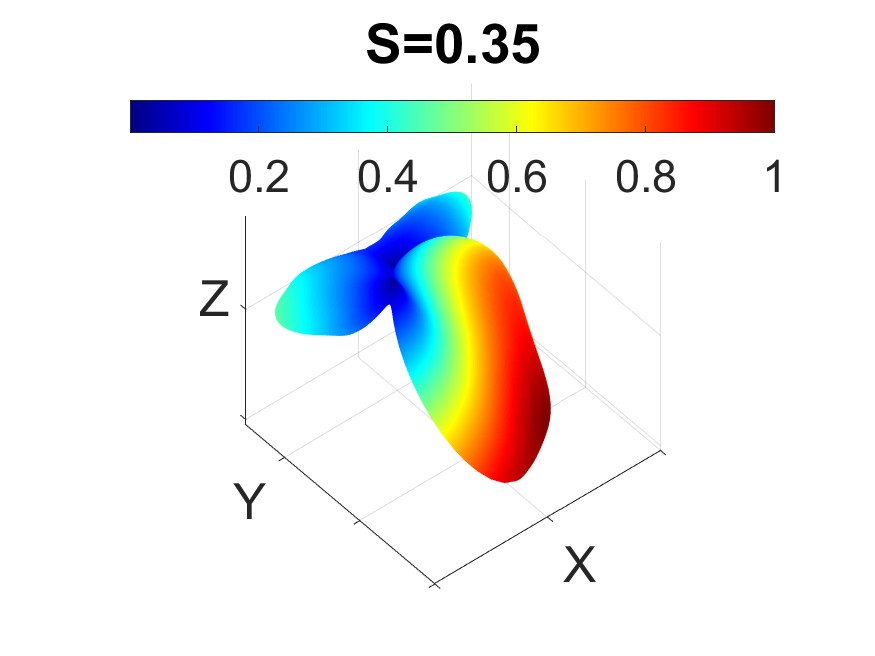}&
    \includegraphics[width=3.5 cm]{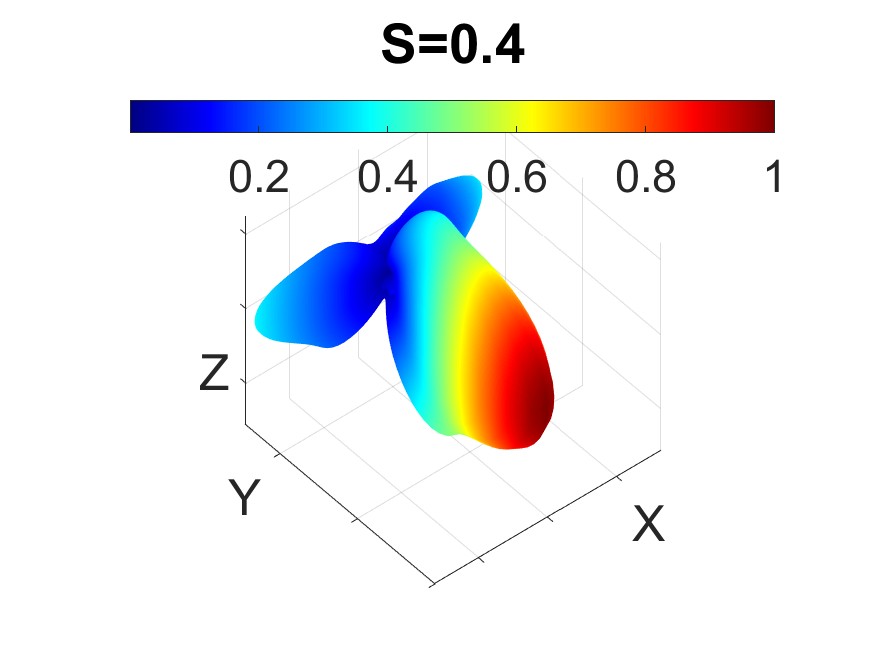}&
    \includegraphics[width=3.5 cm]{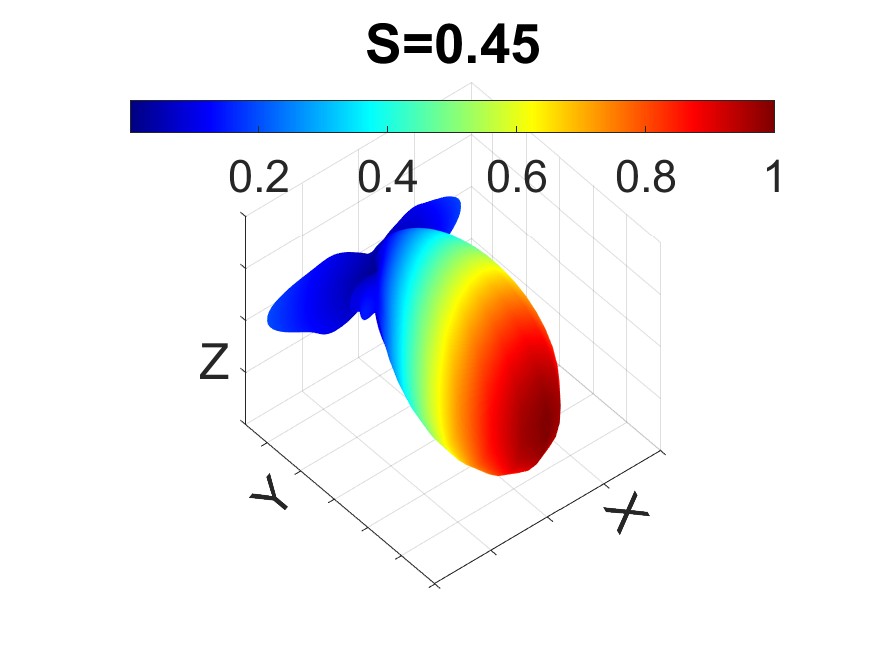}\\
    \includegraphics[width=3.5 cm]{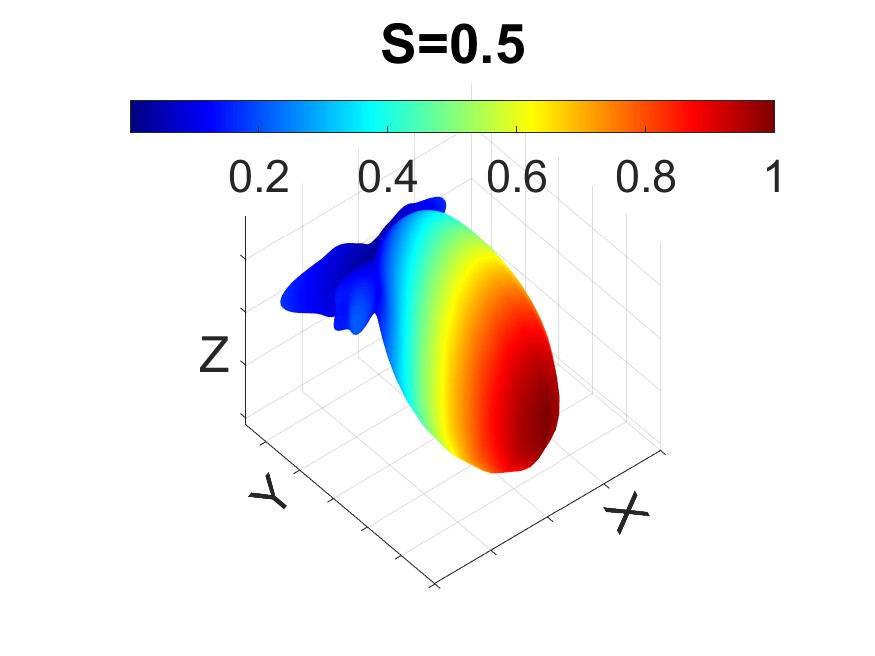}&
    \includegraphics[width=3.5 cm]{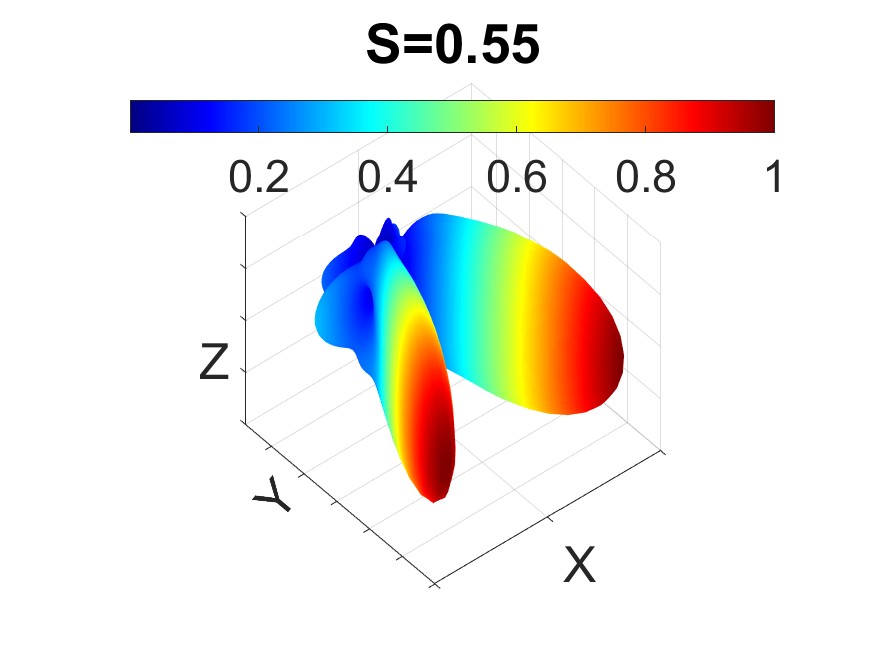}&
    \includegraphics[width=3.5 cm]{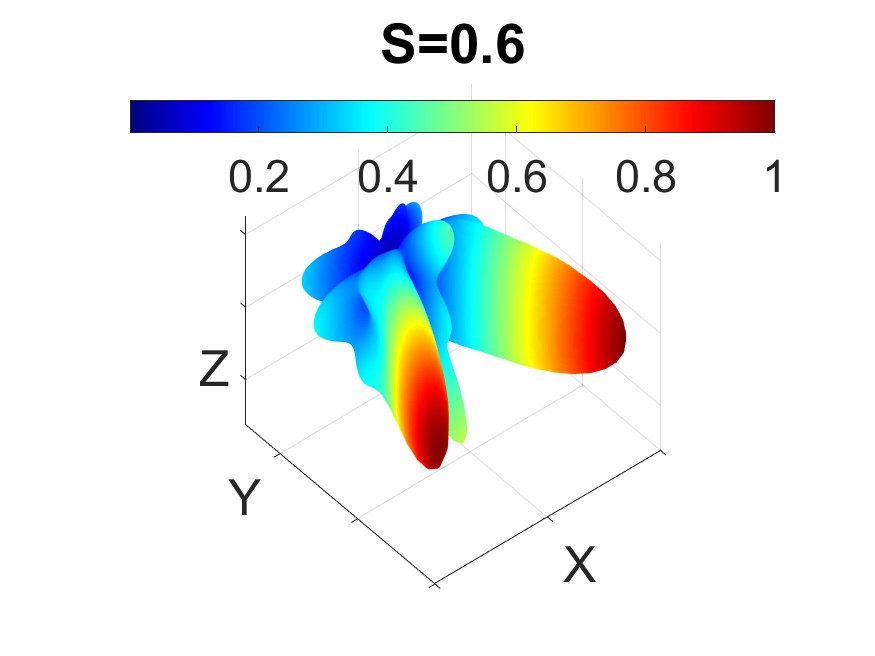}\\
    \includegraphics[width=3.5 cm]{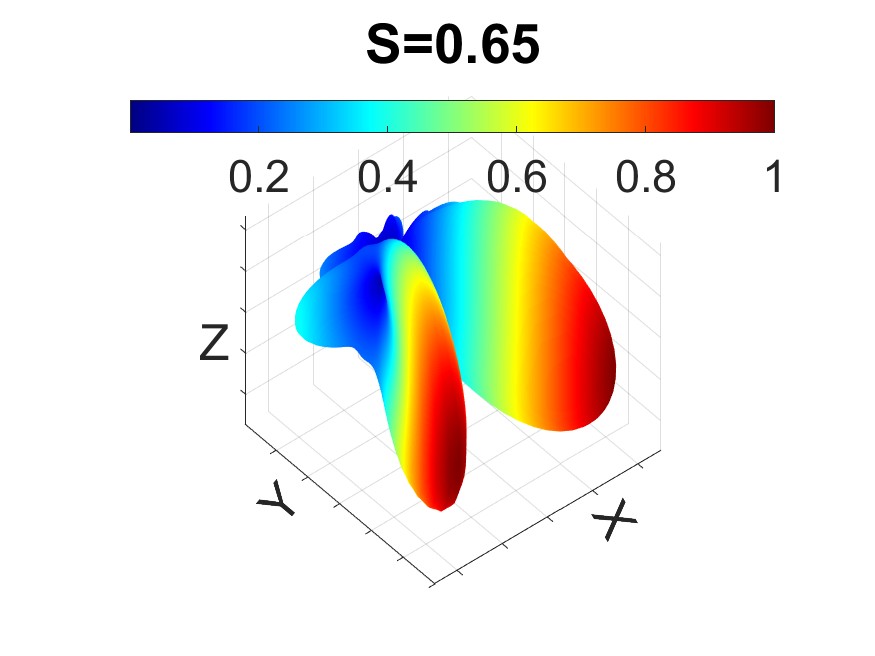}&
    \includegraphics[width=3.5 cm]{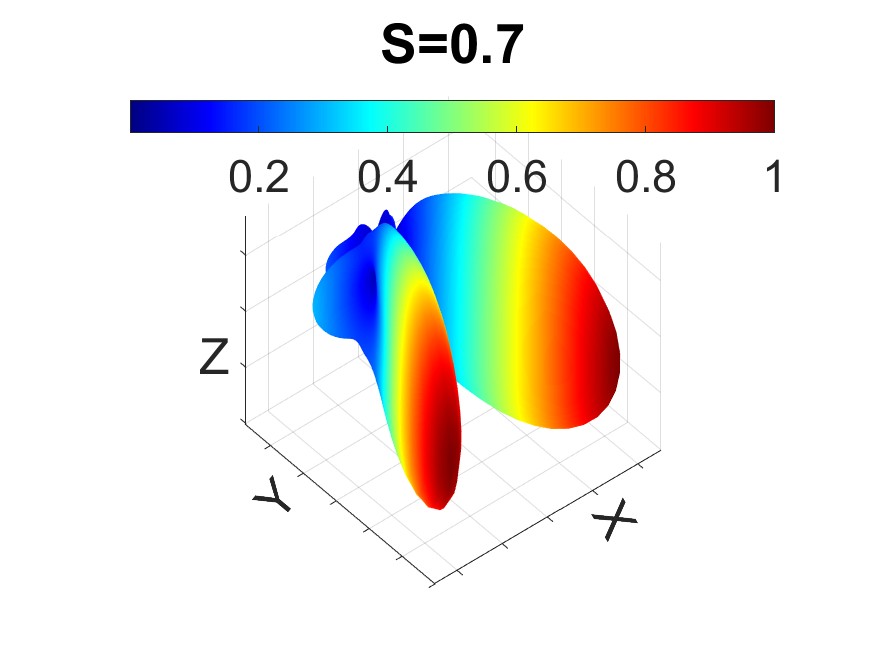}&
    \includegraphics[width=3.5 cm]{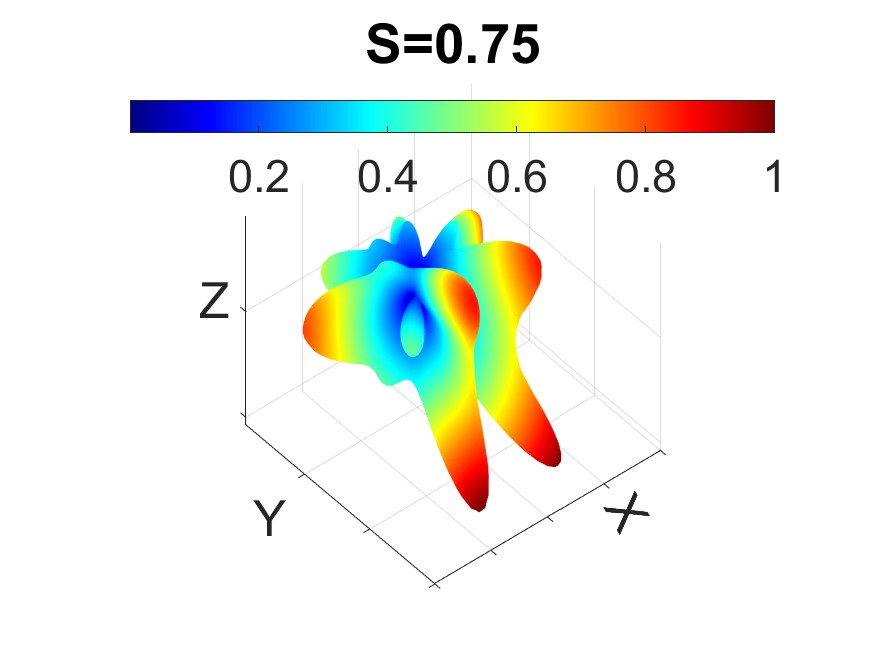}

    \end{tabular}
\caption{Radiation patterns of silver CR with $W=10$ nm for values of $S$ between $S=0.2$ and $S=0.75$.\label{fig::DAgW10}}
\end{figure}

\begin{figure}[H]    
    \begin{tabular}{cccc}
    \includegraphics[width=3.5 cm]{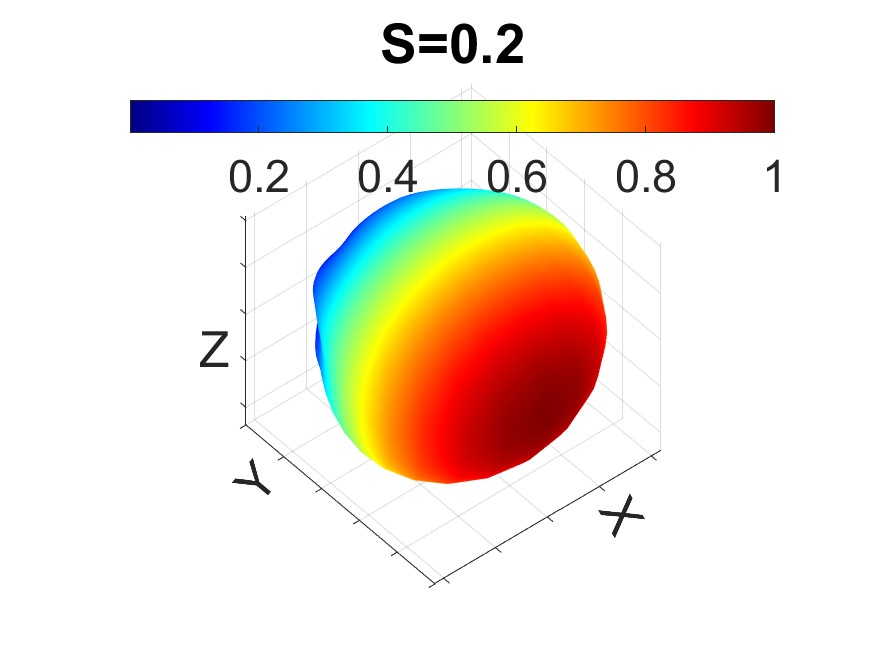}&
    \includegraphics[width=3.5 cm]{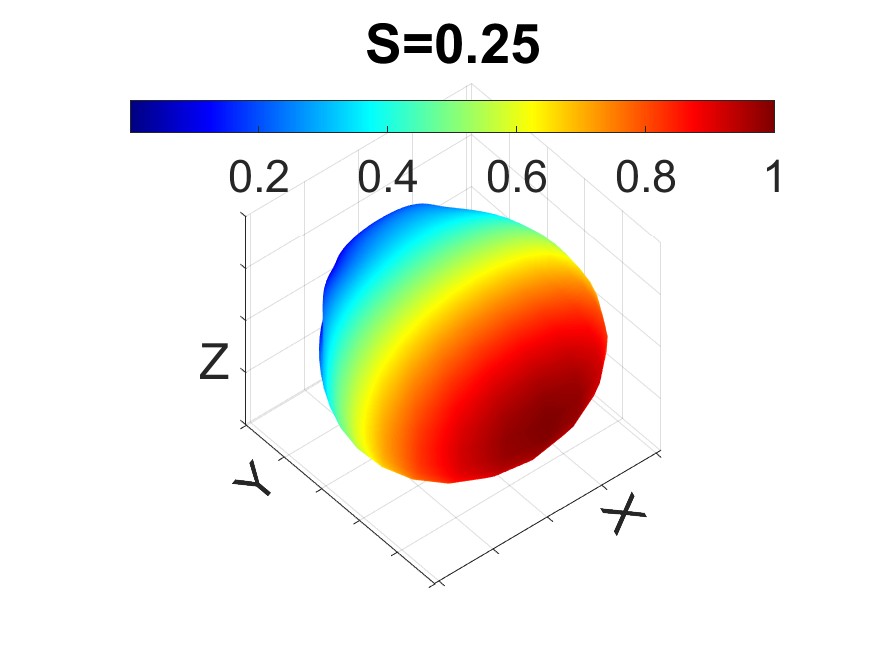}&
    \includegraphics[width=3.5 cm]{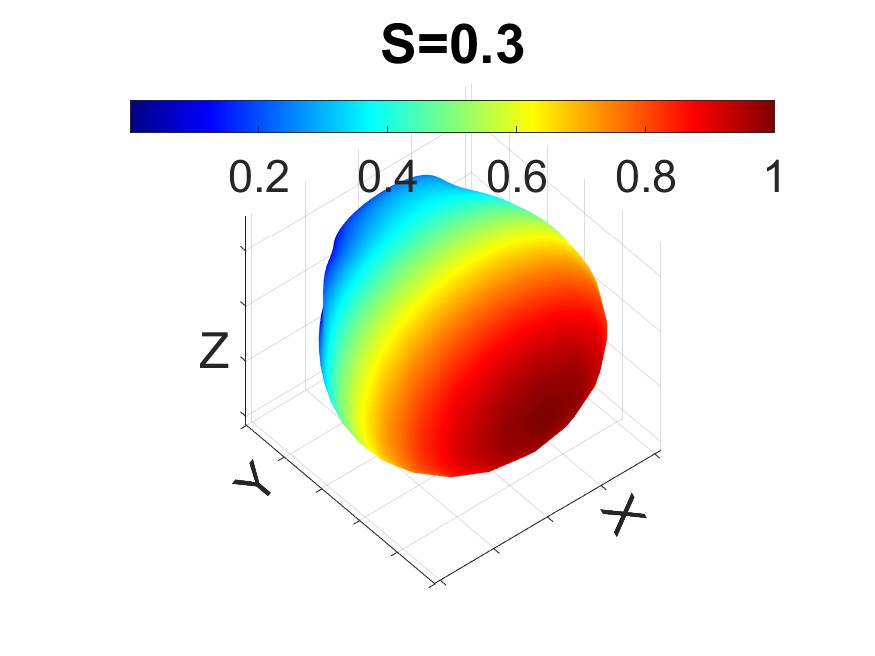}\\
    \includegraphics[width=3.5 cm]{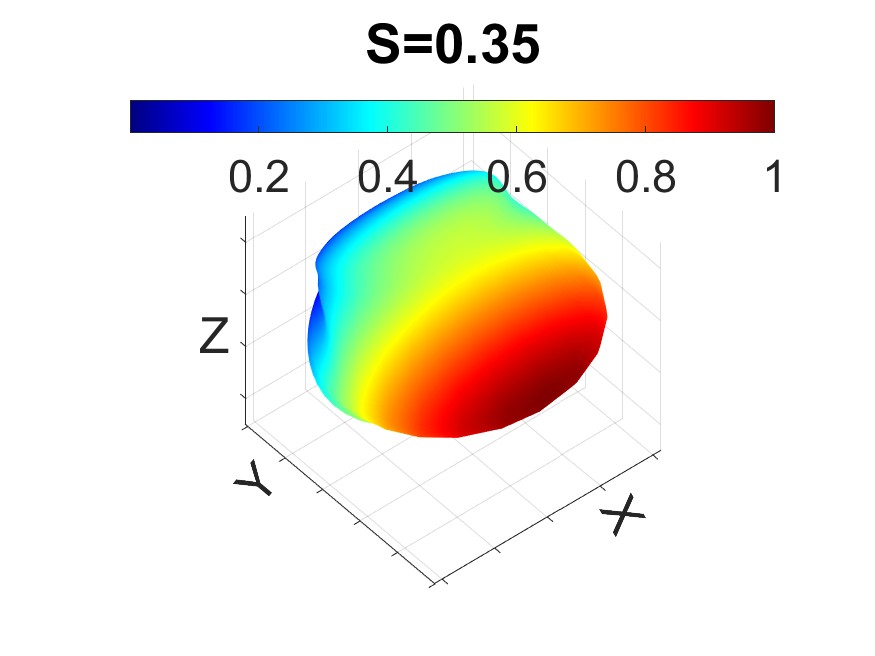}&
    \includegraphics[width=3.5 cm]{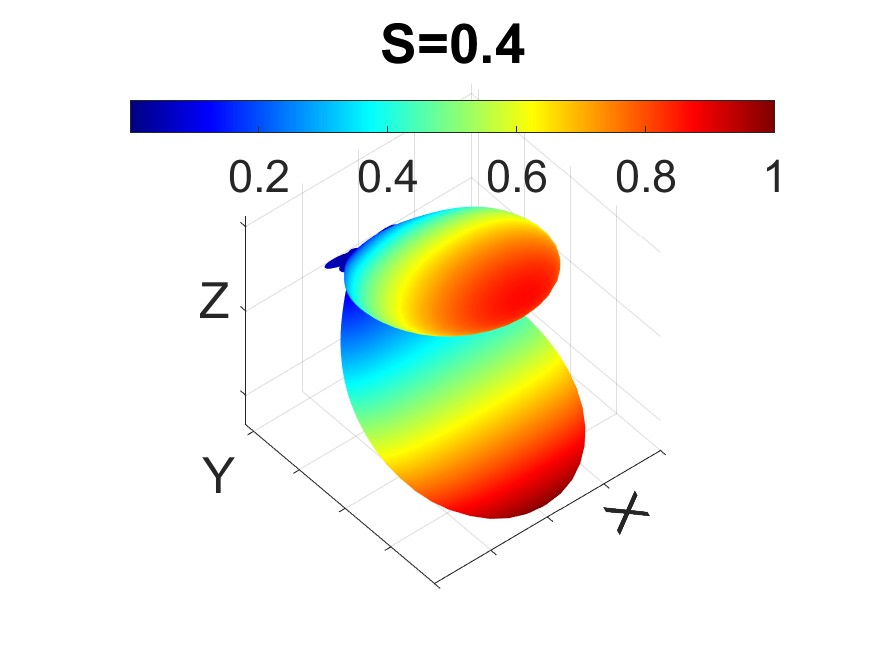}&
    \includegraphics[width=3.5 cm]{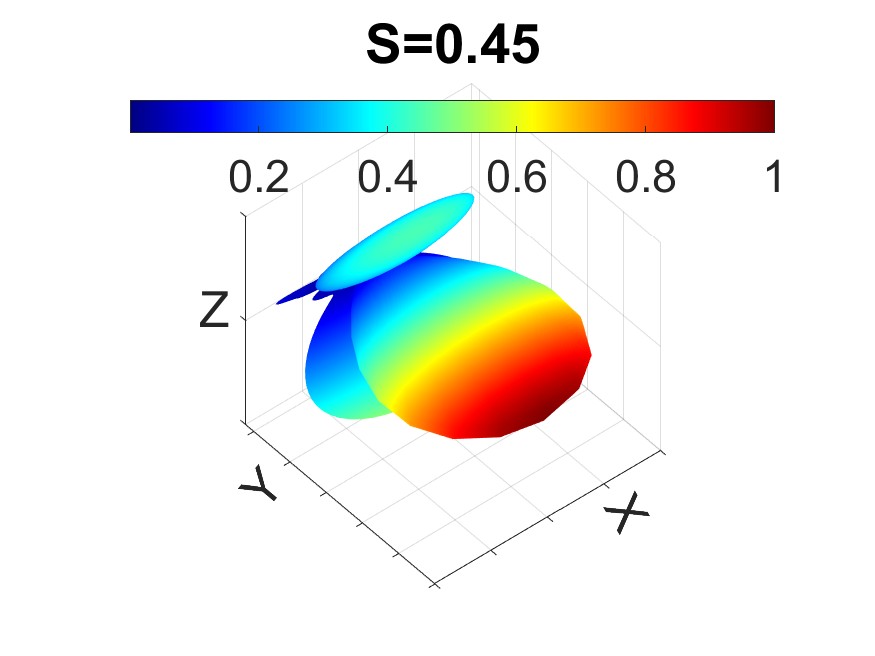}\\
    \includegraphics[width=3.5 cm]{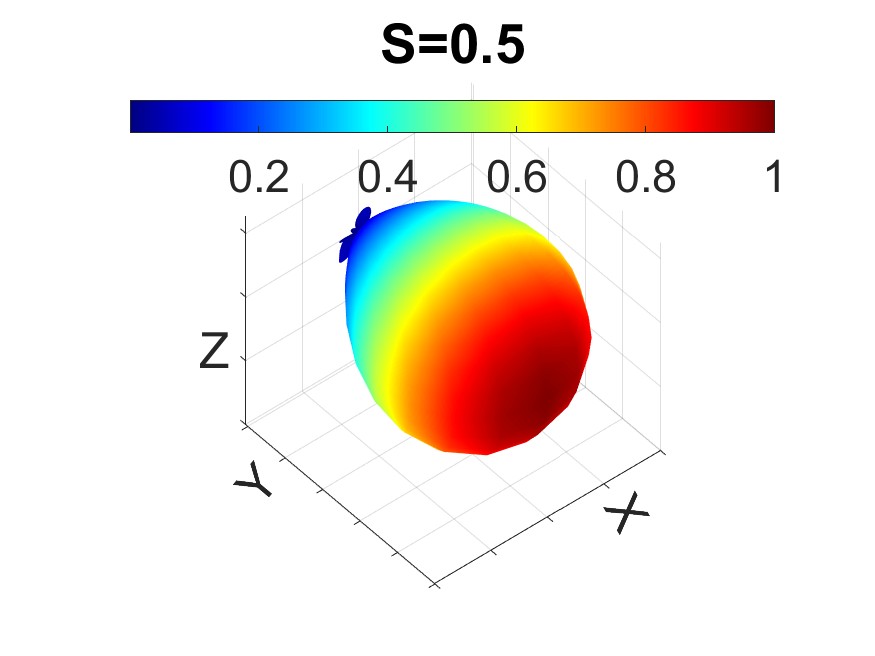}&
    \includegraphics[width=3.5 cm]{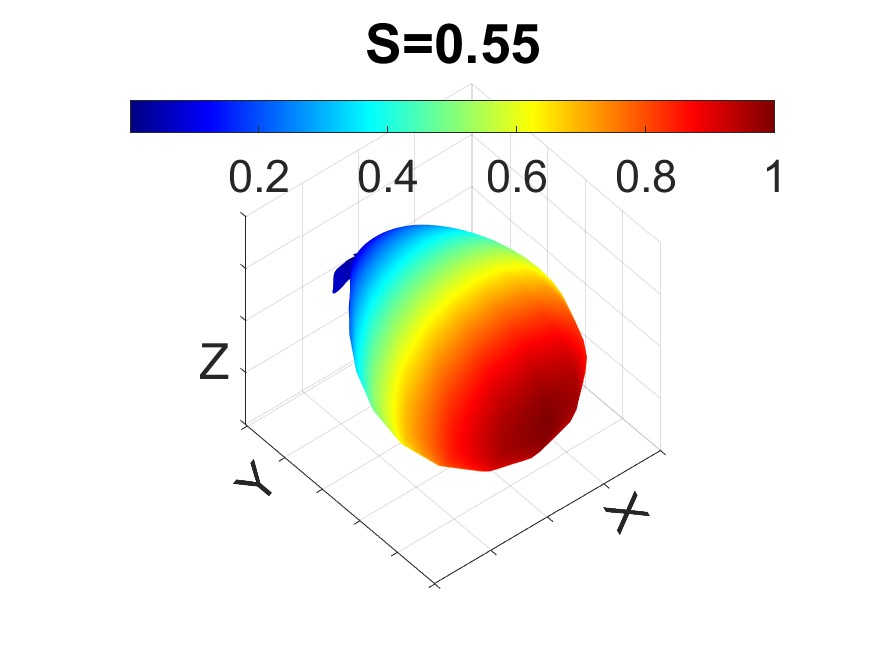}&
    \includegraphics[width=3.5 cm]{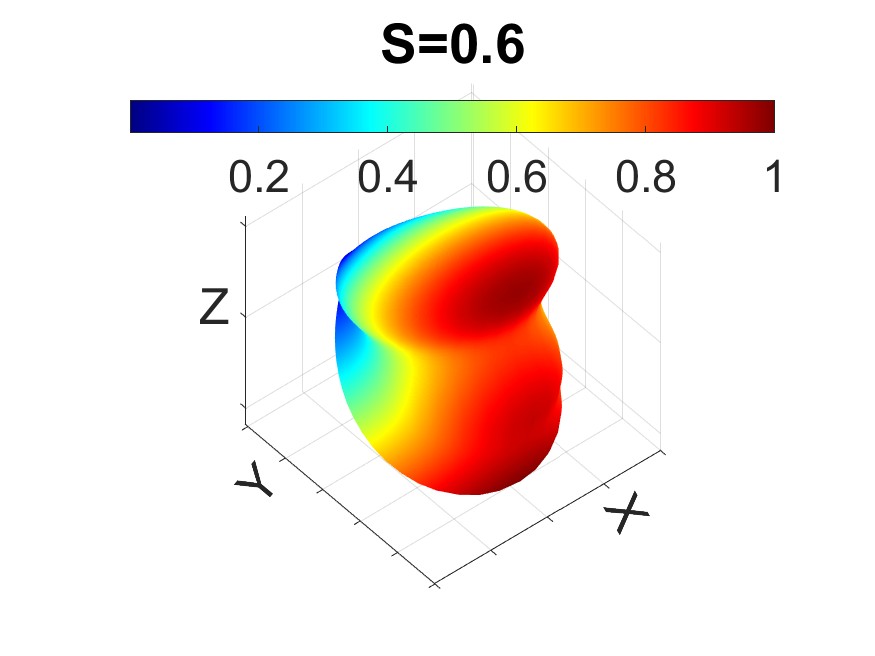}\\
    \includegraphics[width=3.5 cm]{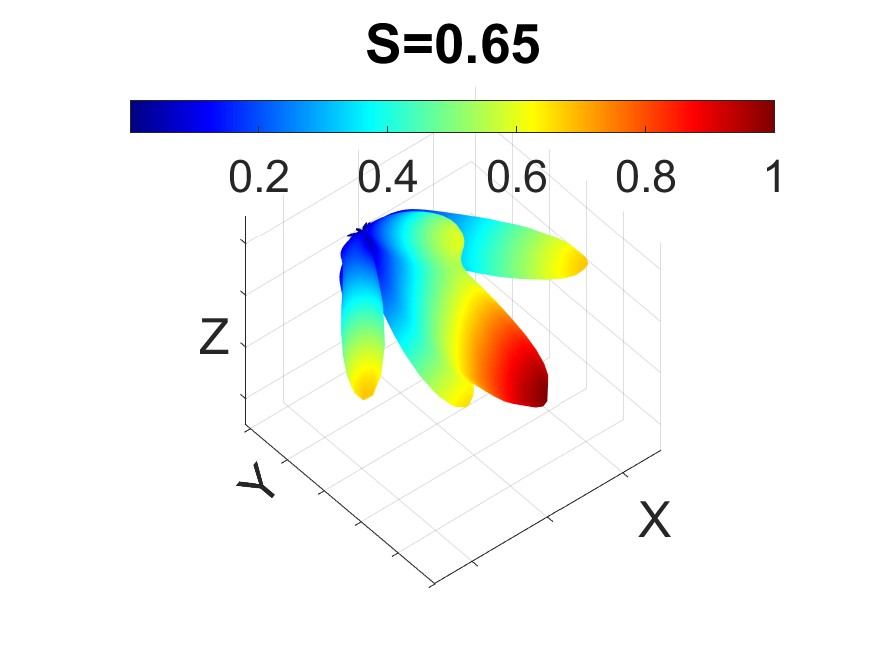}&
    \includegraphics[width=3.5 cm]{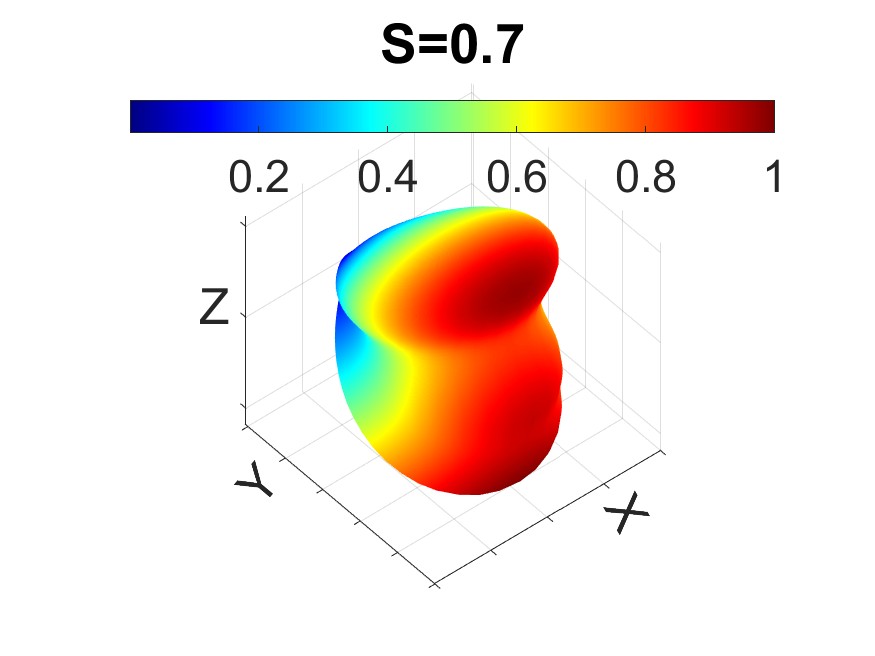}&
    \includegraphics[width=3.5 cm]{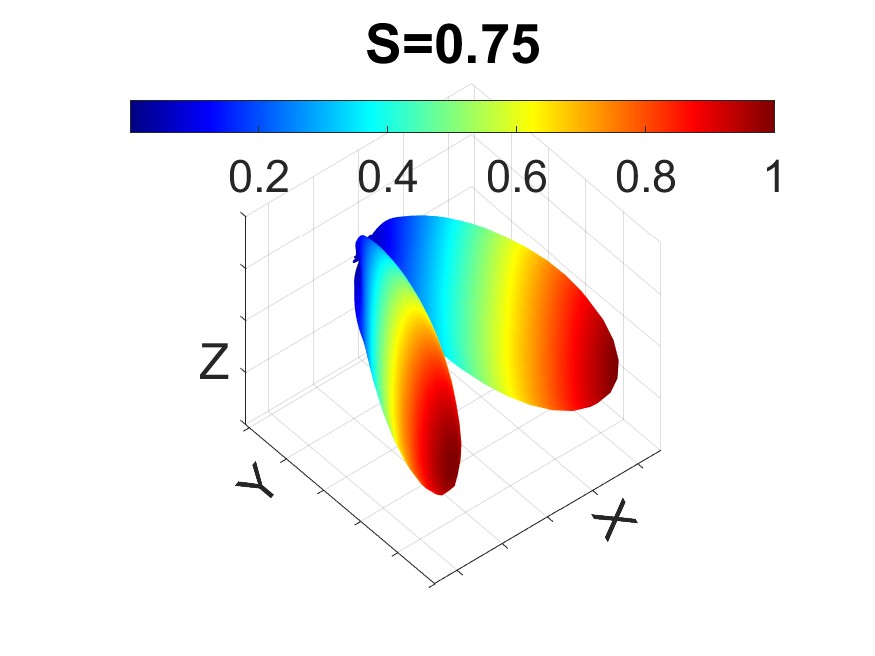}
    \end{tabular}
\caption{Radiation patterns of silver CR with $W=40$ nm for values of $S$ between $S=0.2$ and $S=0.75$.\label{fig::DAgW40}}
\end{figure}

\begin{figure}[H]
  \begin{tabular}{cccc}
    \includegraphics[width=3.5 cm]{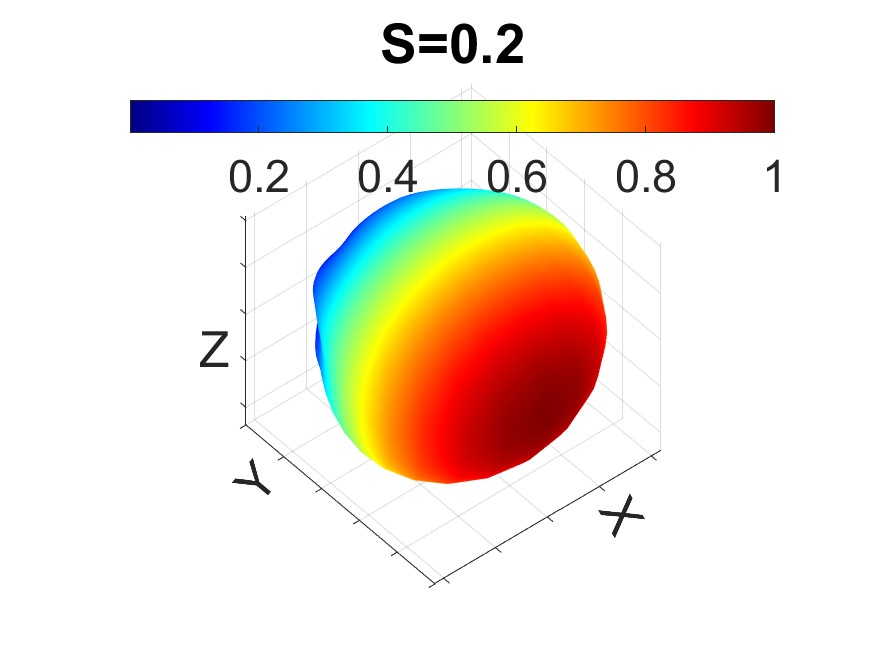}&
    \includegraphics[width=3.5 cm]{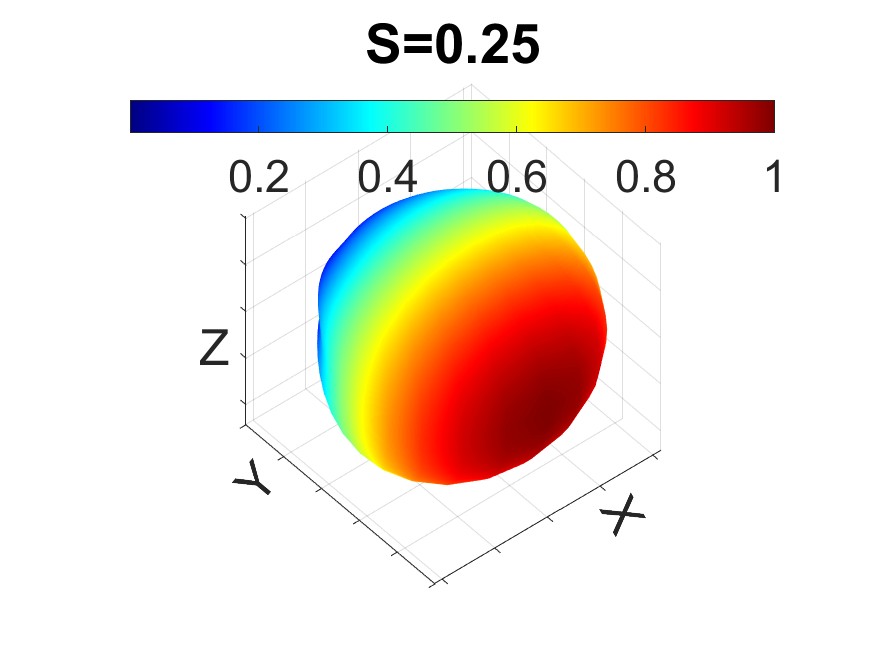}&
    \includegraphics[width=3.5 cm]{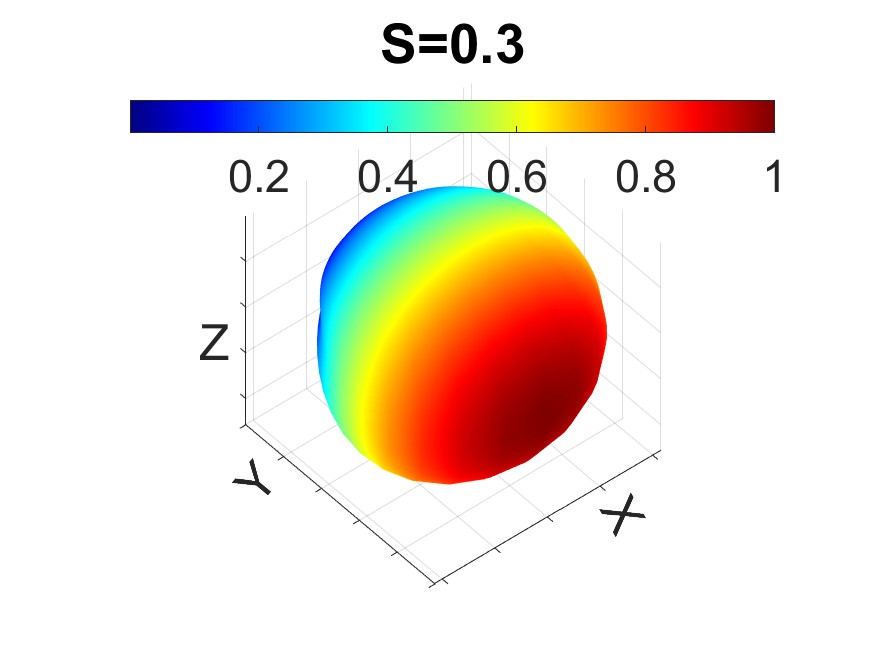}\\
    \includegraphics[width=3.5 cm]{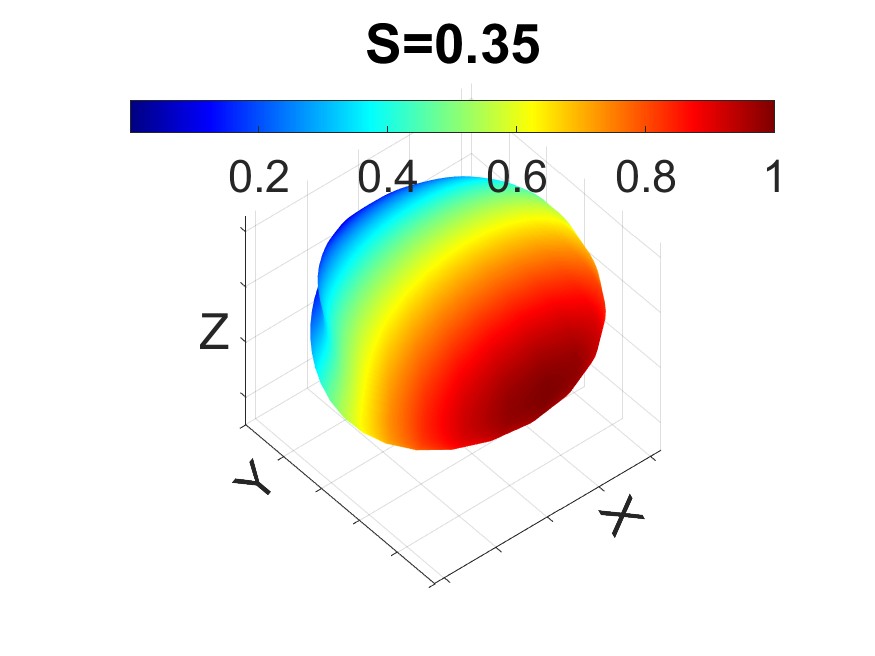}&
    \includegraphics[width=3.5 cm]{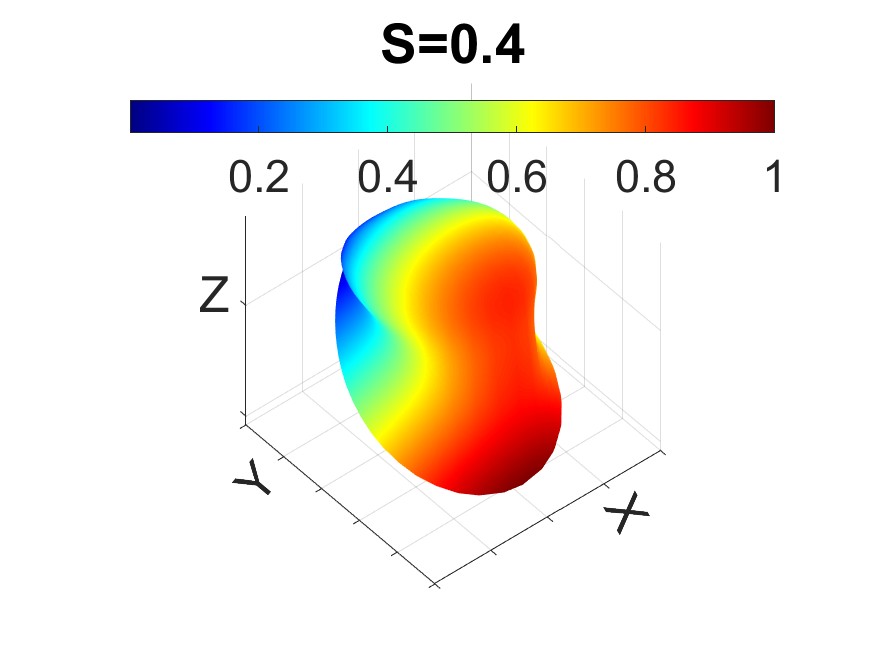}&
    \includegraphics[width=3.5 cm]{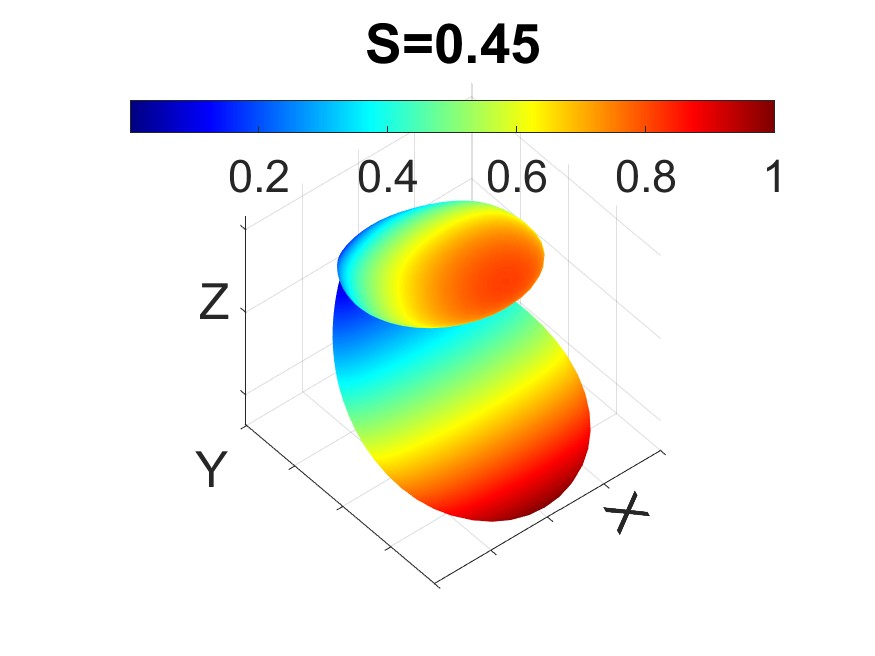}\\
    \includegraphics[width=3.5 cm]{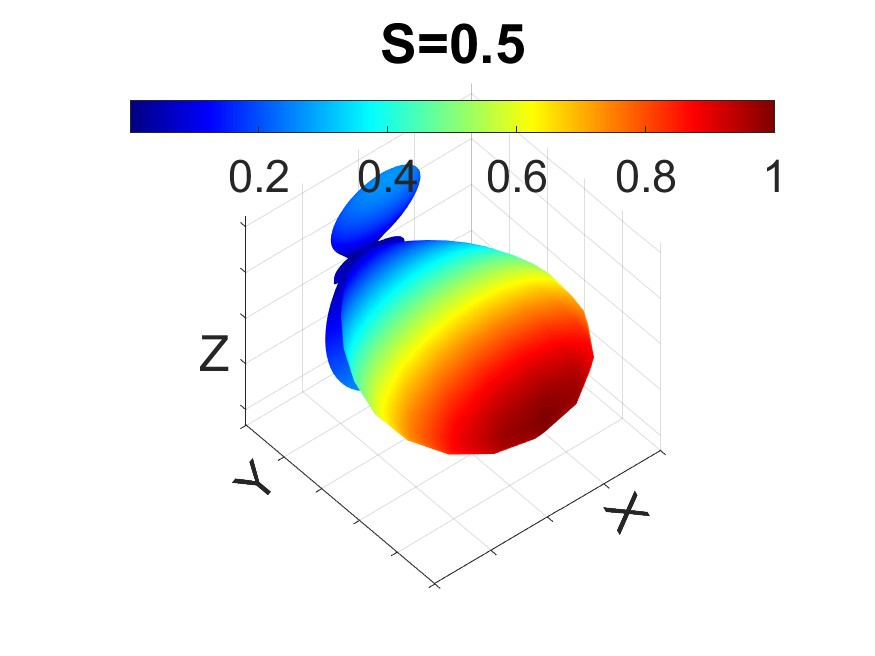}&
    \includegraphics[width=3.5 cm]{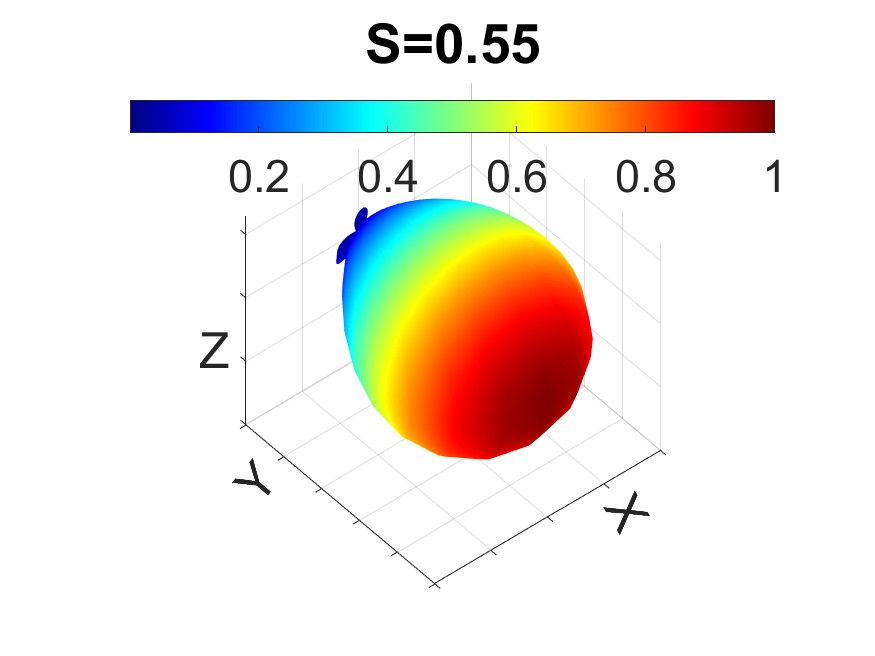}&
    \includegraphics[width=3.5 cm]{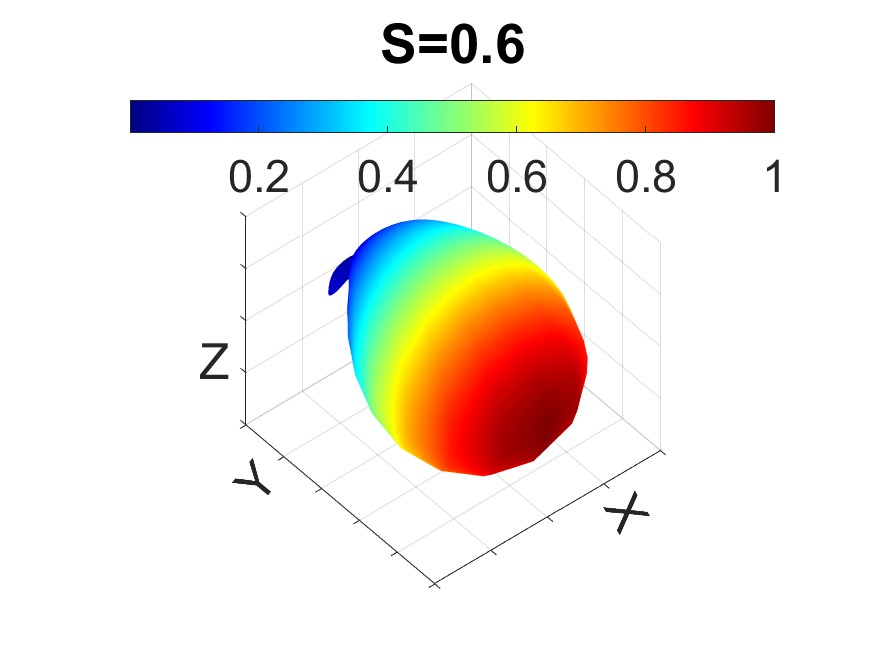}\\
    \includegraphics[width=3.5 cm]{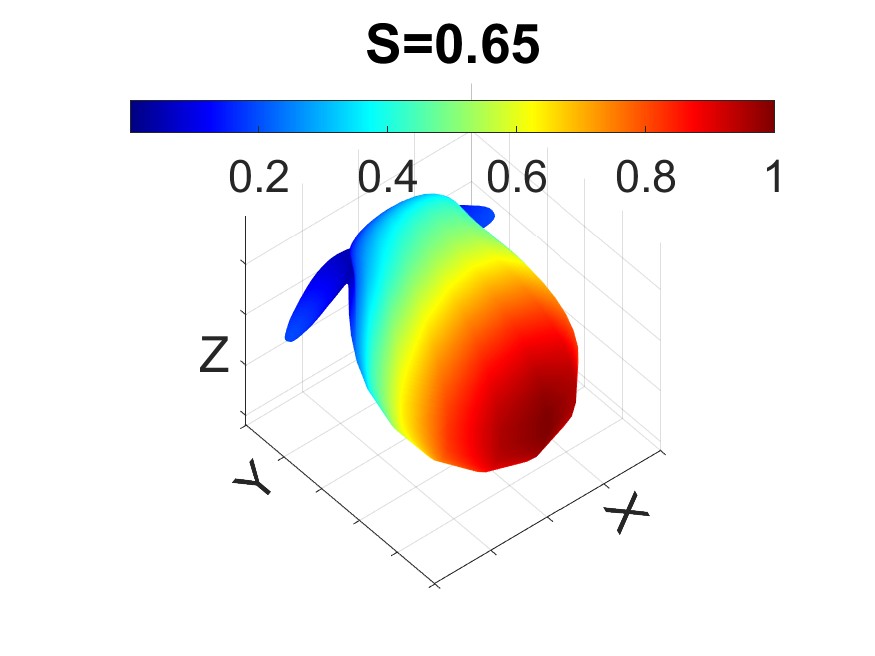}&
    \includegraphics[width=3.5 cm]{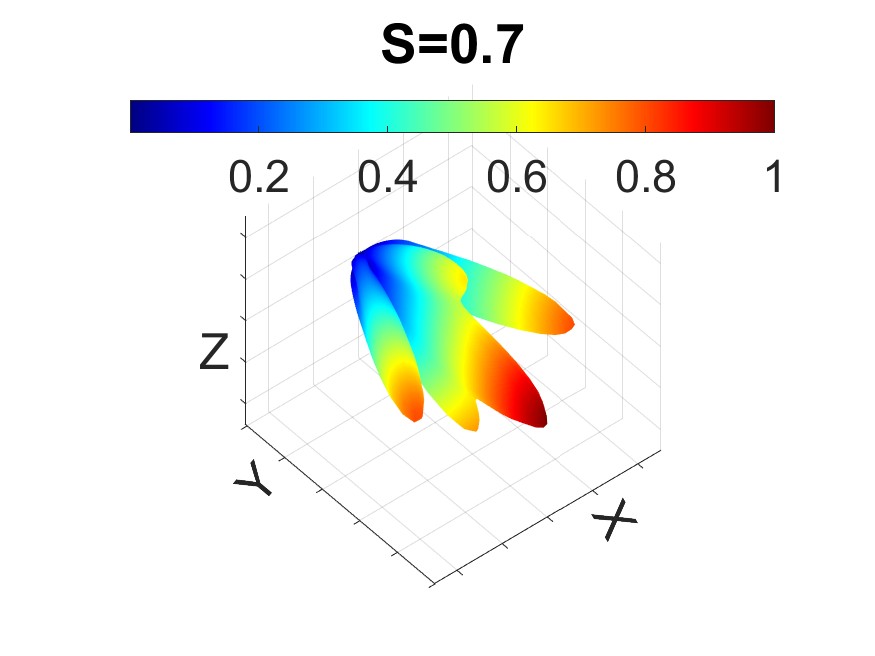}&
    \includegraphics[width=3.5 cm]{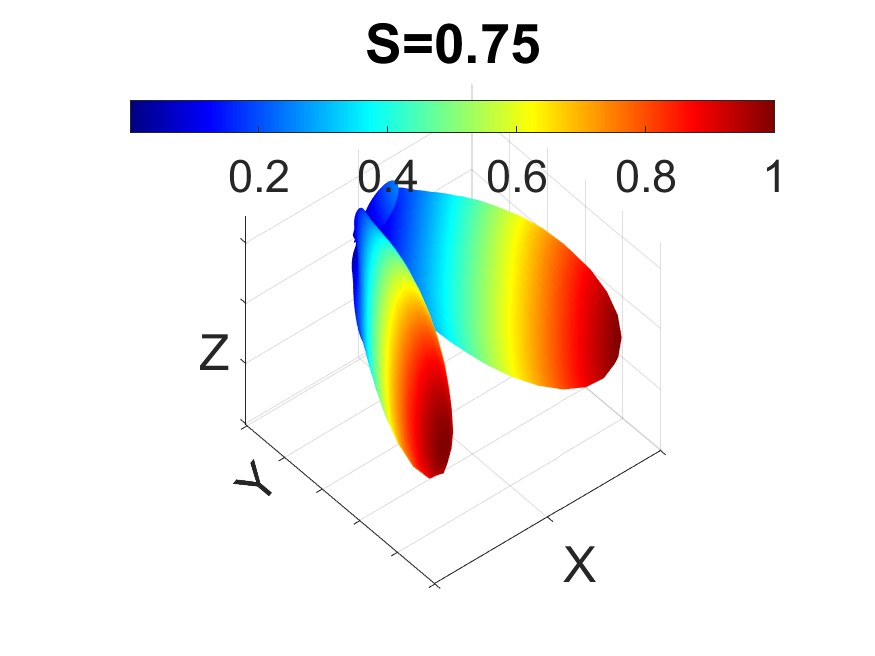}
    \end{tabular}
\caption{Radiation patterns of silver CR with $W=80$ nm for values of $S$ between $S=0.2$ and $S=0.75$.\label{fig::DAgW80}}
\end{figure}

A better understanding of the variations observed in the curves of $D$ can be achieved through a detailed analysis of the radiation patterns in each case.  Such plots are presented for silver CRs with $W=10$ nm, $W=40$ nm, and $W=80$ nm,  respectively, in Figures \ref{fig::DAgW10}--\ref{fig::DAgW80}.

One of the main differences observed at small film thickness ($W=10$ nm) was the presence of relevant amplitude side lobes at the lower values of $S$.  This phenomenon led to reduced directivity values, despite the main lobe exhibiting an narrower beamwidth at $W=10$ nm.  Conversely, the patterns seen at $W=40$ nm and $W=80$ nm were qualitatively similar, indicating a convergence as the thickness $W$ increased.

For thicker-sheet CRs, the gradual narrowing (or broadening) of the on-axis lobe contributed to regions of increasing (or decreasing) directivity values.  The zero observed in the directivity curves arose from the splitting of the main lobe at $S=\lambda_M$  that produced a radiation null along the $Y$ axis.  In contrast, for $W=10$, the local minima in the directivity plots were correlated with the increasing relative weight of the side lobes in the radiation pattern.  As $S$ increased, separation of the main lobe was also noted; however, it occurred along the horizontal plane, unlike the vertical splitting seen at $W=40$ nm and $W=80$ nm.  As $S$ continued to grow, additional lobe bifurcations became apparent, leading to the emergence of a fractal structure in the radiation pattern for very large values of $S$ in extensive CRs, as was analyzed in \cite{chamorro}.

The sequences of radiation diagrams presented in Figures \ref{fig::DAgW10}--\ref{fig::DAgW80} ended up being very similar for the corresponding ones of gold CRs when the values of $S$ were normalized by the respective operation wavelength $\lambda_M$. 

\begin{figure}[H]    
    \begin{tabular}{cc}
    {(a)}&{(b)}\\
    \includegraphics[width=6 cm]{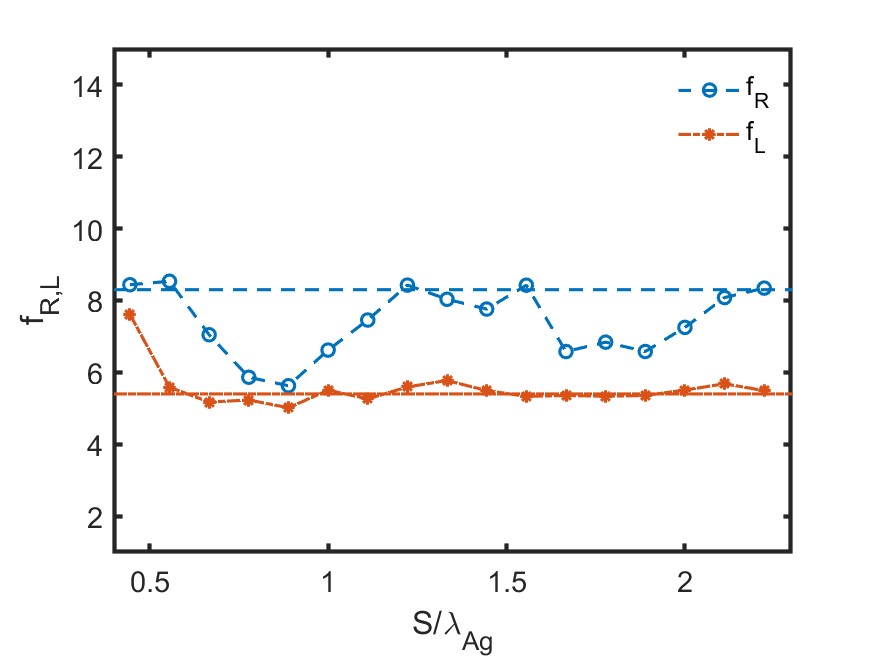}&
    \includegraphics[width=6 cm]{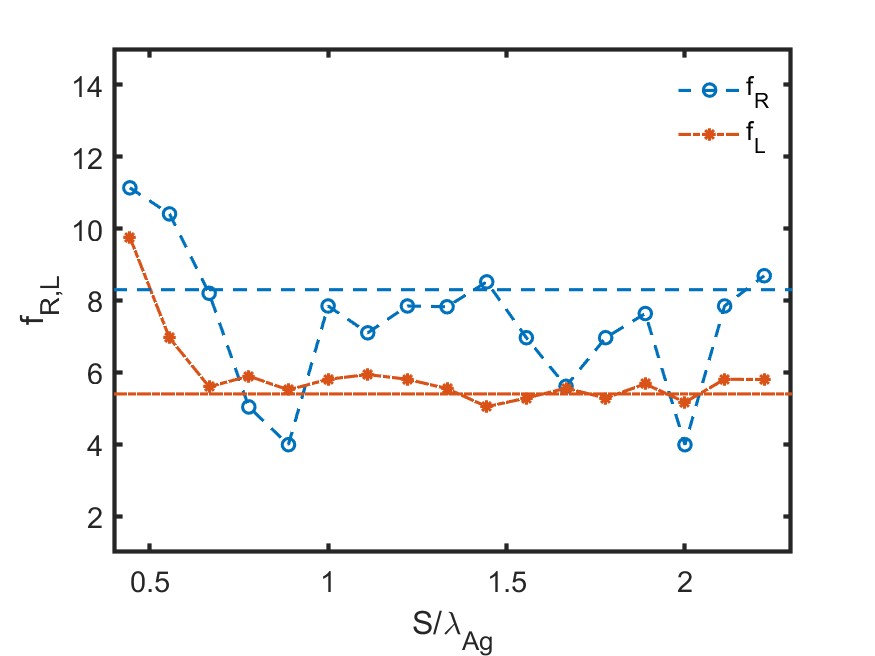}\\
     {(c)}&{(d)}\\
    \includegraphics[width=6 cm]{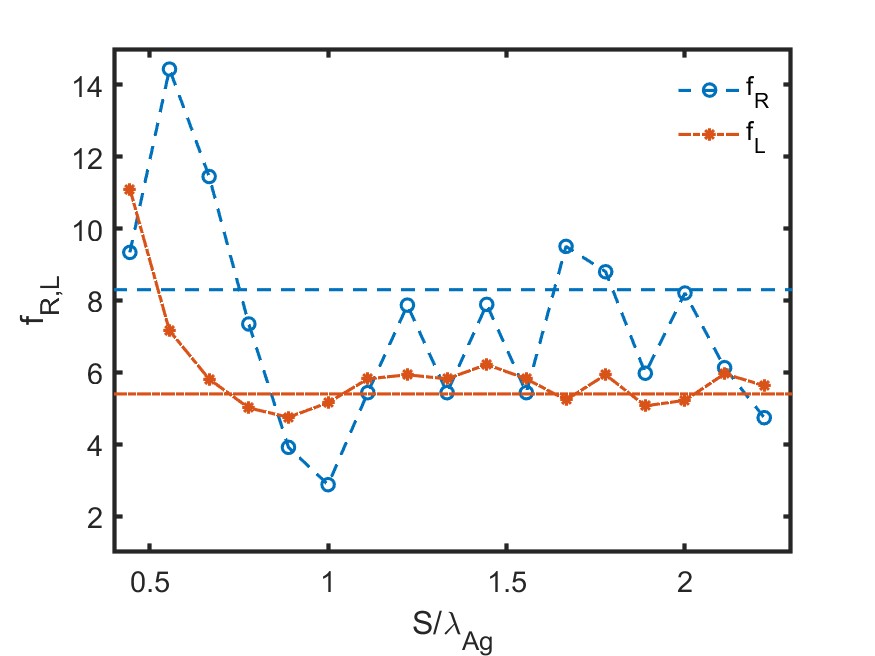}&
    \includegraphics[width=6 cm]{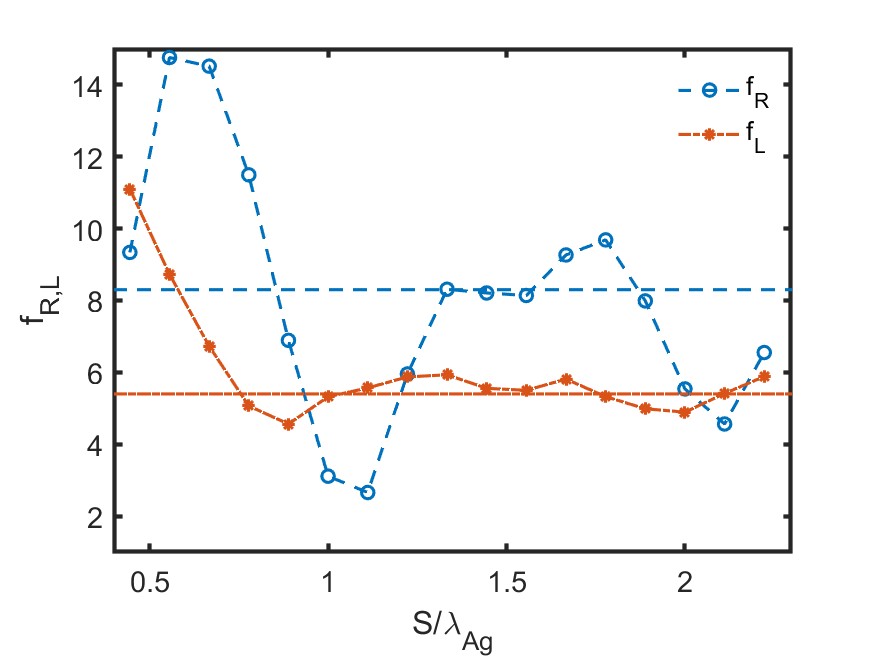}\\
    \end{tabular}
\caption{Purcell, $f_R$, and normalized loss, $f_L$, factors of silver CR as a function of the normalized feeder position $S/\lambda_M$ for four different reflector film thicknesses: (\textbf{a}) $W=10$ nm, (\textbf{b}) $W=20$ nm, \mbox{(\textbf{c}) $W=40$ nm}, and (\textbf{d}) $W=80$ nm.  The horizontal lines describe the corresponding reference levels without the CR.\label{fig::cornerplata}}
\end{figure}

\begin{figure}[H]   
    \begin{tabular}{cc}
    {(a)}&{(b)}\\
    \includegraphics[width=6 cm]{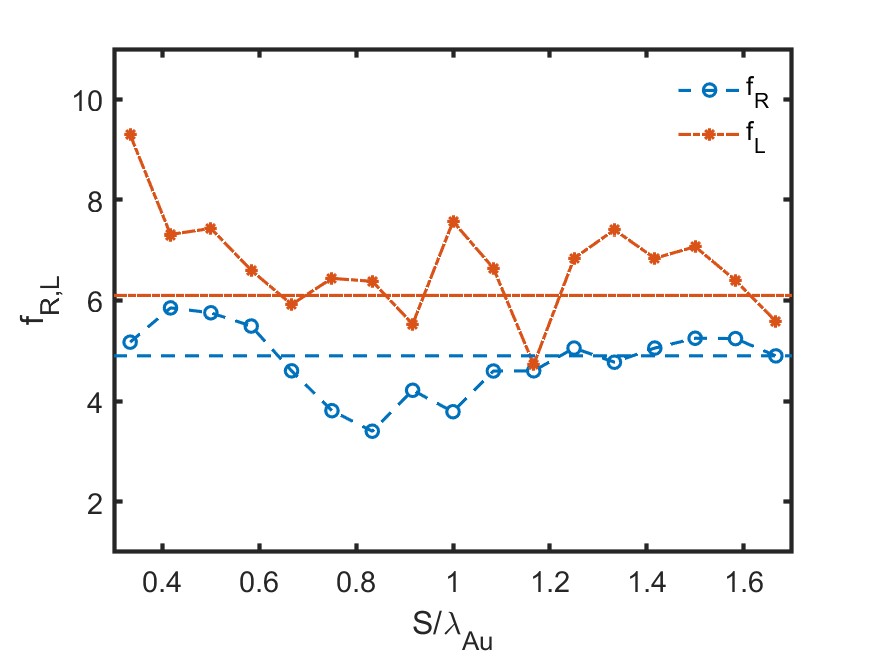}&
    \includegraphics[width=6 cm]{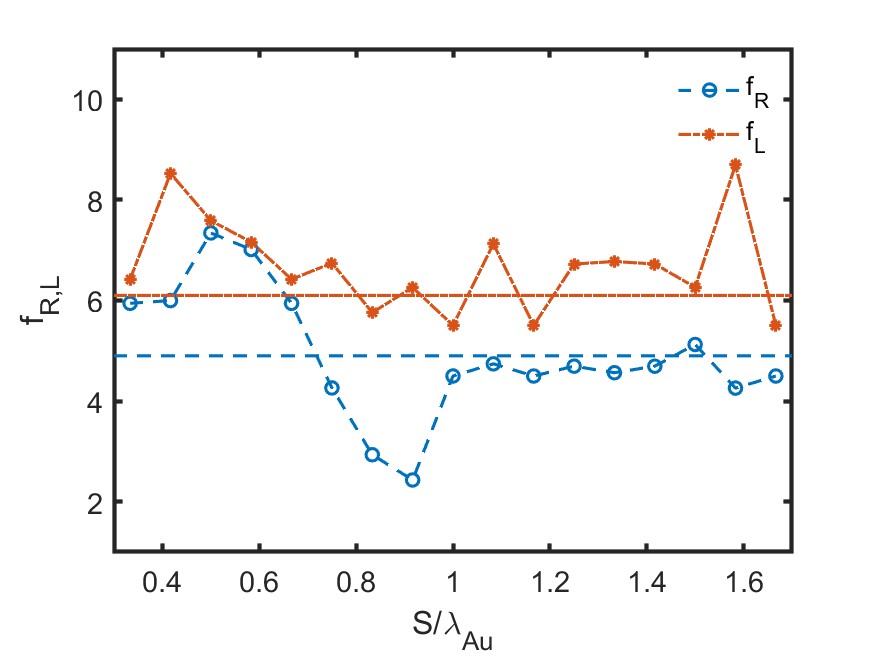}\\
     {(c)}&{(d)}\\
    \includegraphics[width=6 cm]{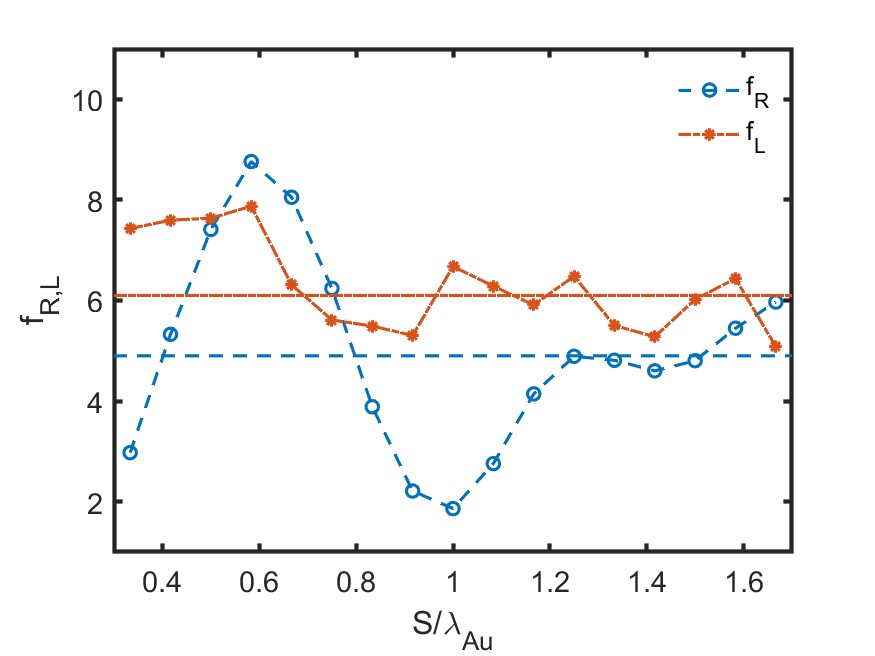}&
    \includegraphics[width=6 cm]{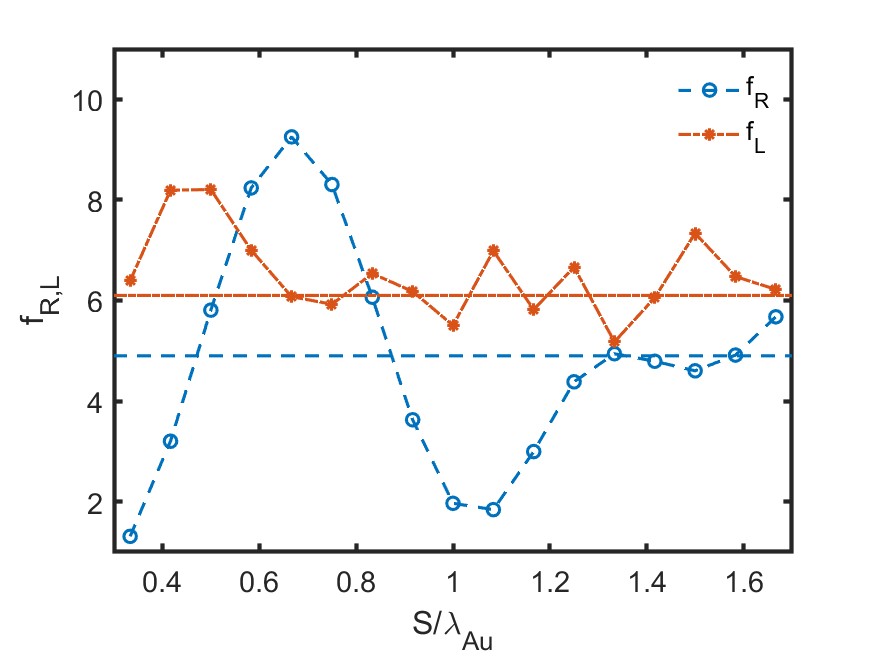}\\
    \end{tabular}
  \caption{Radiative transition rate enhancement, $f_R$, and normalized loss, $f_L$, factors of gold CR as a function of the normalized feeder position $S/\lambda_M$ for four different reflector film thicknesses: \mbox{(\textbf{a}) $W=10$ nm}, (\textbf{b}) $W=20$ nm, (\textbf{c}) $W=40$ nm, and (\textbf{d}) $W=80$ nm.\label{fig::corneroro}}
\end{figure}

We now examine the impact of the corner reflector on the losses and the emission rate of the dipole-NP system.  Figure \ref{fig::cornerplata} displays the $f_R$ and the normalized loss factor $f_L$ of NAs incorporating silver CRs of four different thicknesses, $W$, as a function of the normalized distance, $S/\lambda_{Ag}$,  between the silver NP and the corner apex.  The corresponding results for the gold NAs are shown in Figure \ref{fig::corneroro}.  In all the figures, the horizontal lines pinpoint the reference values for the NPs without the CRs.  
\vspace*{0.5cm}

\unskip

\unskip

\vspace{-12pt}

\unskip

For silver CRs, the normalized loss factors displayed moderate deviations from the reference values except for very small values of $S/\lambda_{Ag}$, where a steep increase in the losses occurred due to stronger interactions with the nearby CR metal films when the thickness, $W$, was sufficiently large.  In contrast, for gold CRs, Figure \ref{fig::corneroro} indicates larger fluctuations in the loss factor around the reference levels.

The calculated values of the Purcell factor, $f_R$, also exhibited a pronounced effect due to the interaction with the corner sheets when $W$ was sufficiently large.  This interaction significantly impacted both silver and gold NAs, and the associated $S/\lambda_{M}$ range was larger than that observed for $f_L$ in silver NRs, increasing with $W$.  The variation in $S/\lambda_M$ followed a distinct pattern: $f_R$ reached a maximum near the first peak of directivity and then dropped to a minimum around the directivity null at $S/\lambda_M=1$ before rising again near the second peak.

\begin{figure}[H]   
    \begin{tabular}{cc}
    {(a)}&{(b)}\\
    \includegraphics[width=6 cm]{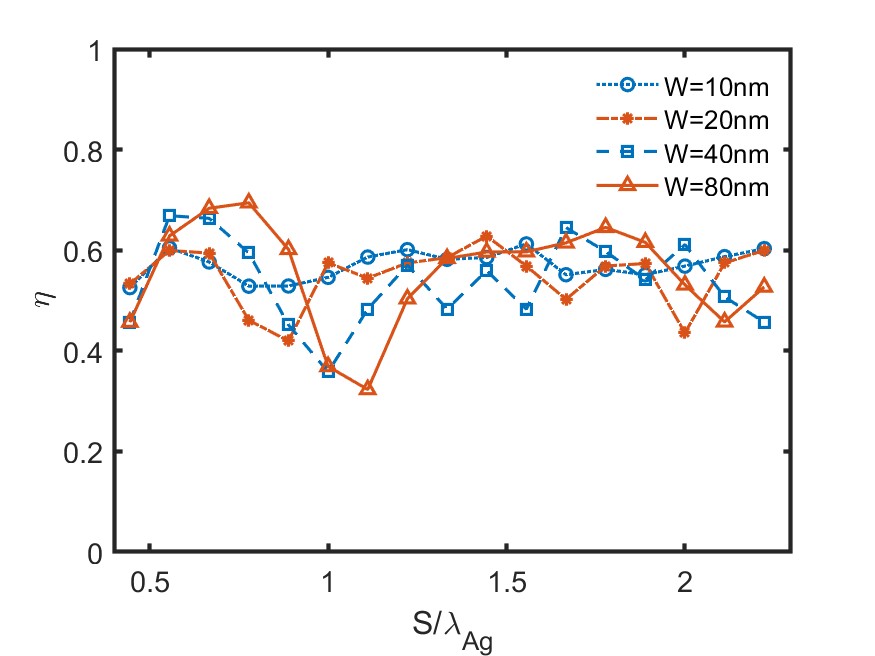}&
    \includegraphics[width=6 cm]{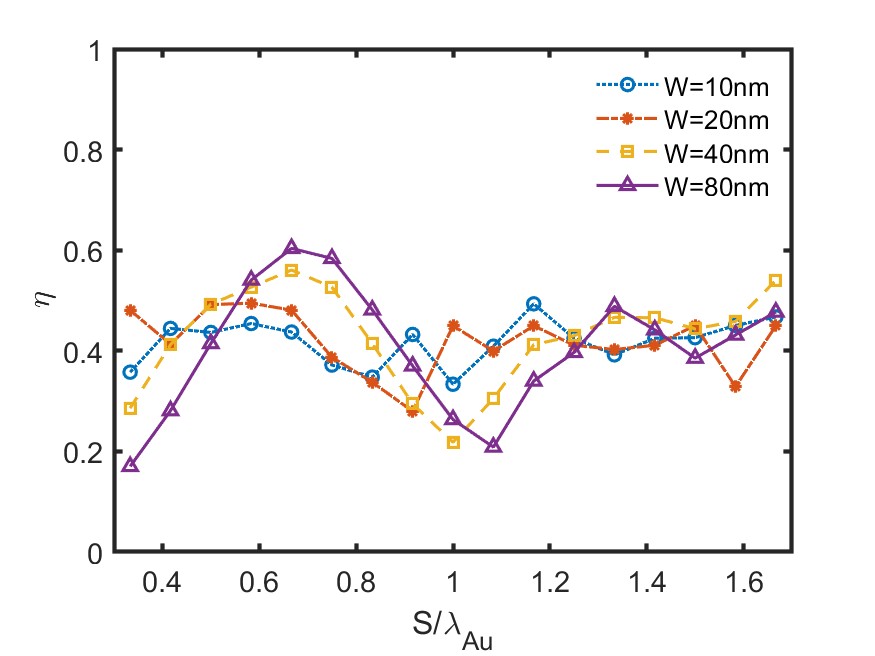}\\
    \end{tabular}
  \caption{Radiation efficiency of silver (\textbf{a}) and gold (\textbf{b}) CR NA as a function of the normalized feeder position $S/\lambda_M$ for four different reflector film thicknesses $W$. \label{fig::eta}}
\end{figure}

Figure \ref{fig::eta} displays the radiation efficiency $\eta$ of the gold and silver CR NAs as a function of the feeder position $S$. As discussed in Section \ref{MM}, this parameter is a very good approximation regarding the quantum efficiency of the NA for a highly efficient fluorophore with a negligible non-radiative transition rate. For thin metal sheets, with $W=10$ nm values, the fluctuations in $\eta$ as a function of $S$ were of small amplitudes. Larger efficiencies were observed for silver NAs, with values close to $\eta=0.6$, compared to gold NAs, which reached values around $\eta=0.4$. As the value of $W$ increased, the amplitude of the fluctuations in $\eta$ with $S$ became larger. Once again, the peak value of $\eta$ for silver, close to $0.7$, was higher than that for gold, which approached $0.6$.

For silver CRs with  $W=80$ nm values, we found that near the first directivity maximum at $S=0.30$ $\upmu$m ($S/\lambda_{Ag}=0.67$), the values were  $D=20.1$, $f_R=14.5$, and $f_L=6.7$.  The second maximum yielded a significantly higher directivity of $D=40.4$ at $S=0.55$ $\upmu$m ($S/\lambda_{Ag}=1.22$), with $f_R=6.0$ and $f_L=5.9$. Thus, the CR provided very large improvements in the directivity value when compared to those of the electric dipole emitter, with $D=1.5$.

Considering the reference values for the NP emitter, $f_R=8.3$ and $f_L=5.4$, the first maximum offered a significant increase in both directivity and radiative emission rate enhancement factor, with only a modest rise in losses.  At the second maximum, the directivity was much larger, though there was a slight reduction in $f_R$ in relation to the reference.  The loss factor, however, remained close to that of the emitter without the CR.  The gain figure of merit values were $G=13.8$ and $G=20.4$ for the first and second maxima, respectively, largely exceeding the value of $G=0.9$ obtained for the NA without the CR.  

For the gold NA, the reference values were $f_R=4.9$ and $f_L=6.1$.  With a $W=80$ nm thick.  CR NA, at the first maximum, located at $S=0.35$ $\upmu$m ($S/\lambda_{Au}$=0.58), the directivity was $D=18.0$, with $f_R=8.2$ and $f_L=7.0$.  The second maximum occurred at $S=0.75$ $\upmu$m ($S/\lambda_{Au}=0.58$), where $D=29.0$, $f_R=4.4$, and $f_L=6.7$.

For the gold CRs, the first maximum provided a high directivity and an improvement in the $f_R$ value. At the second maximum, while the directivity was even greater, the $f_R$ was close to the reference value.  In both cases, the losses were similar to those of the gold NP without the CR.  The gain figure of merit values were $G=8.0$ and $G=11.5$ for the first and second maxima, respectively, significantly exceeding the value of $G=0.7$ obtained for the NA without the CR but significantly in value lower than the values obtained for silver CR NAs.  In addition to the impact of the better figures for the $f_R$ and $f_L$ values in the case of silver NAs, the lower values of $G$ obtained in the case of gold CR NAs were also due to the fact that $L/\lambda_{M}$ was smaller in the case of gold.

\begin{figure}[H] 
    \begin{tabular}{cc}
    {(a)}&{(b)}\\
    \includegraphics[width=6 cm]{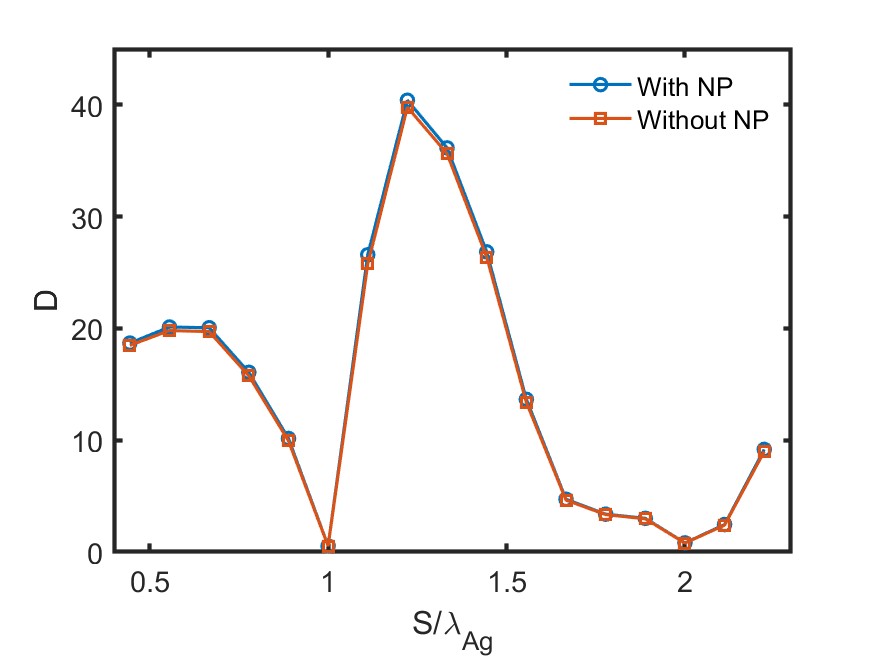}&
    \includegraphics[width=6 cm]{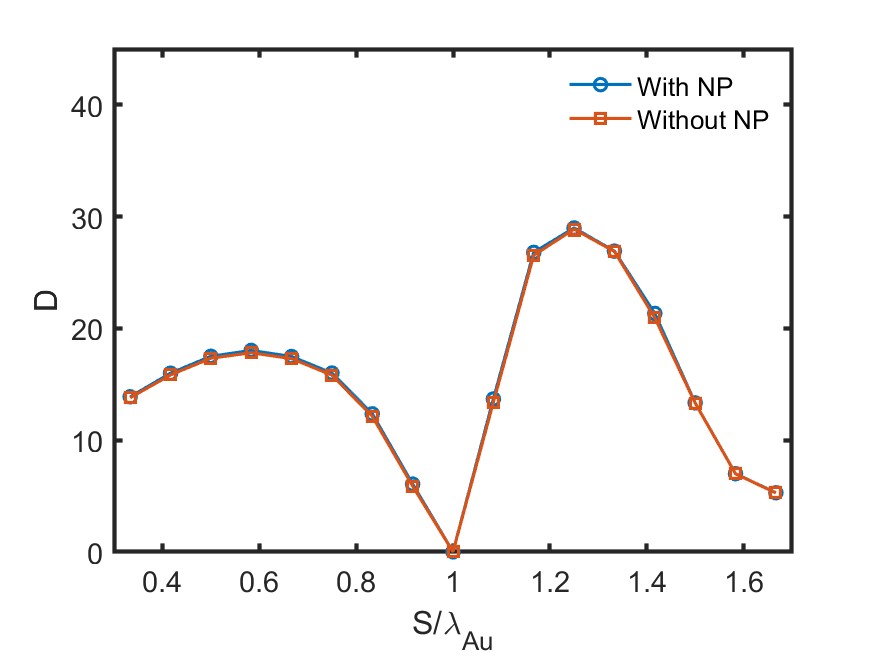}\\
    {(c)}&{(d)}\\
    \includegraphics[width=6 cm]{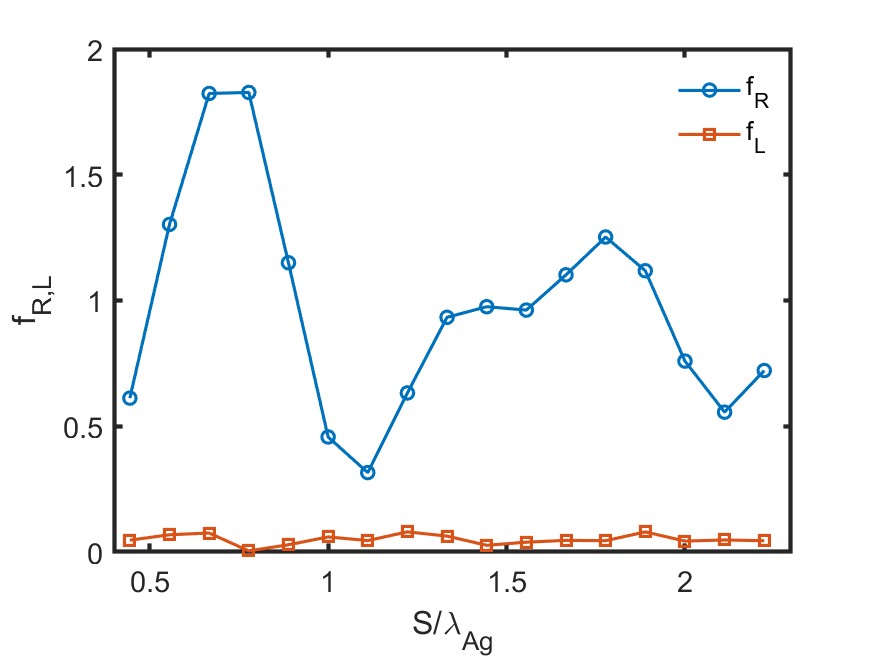}&
    \includegraphics[width=6 cm]{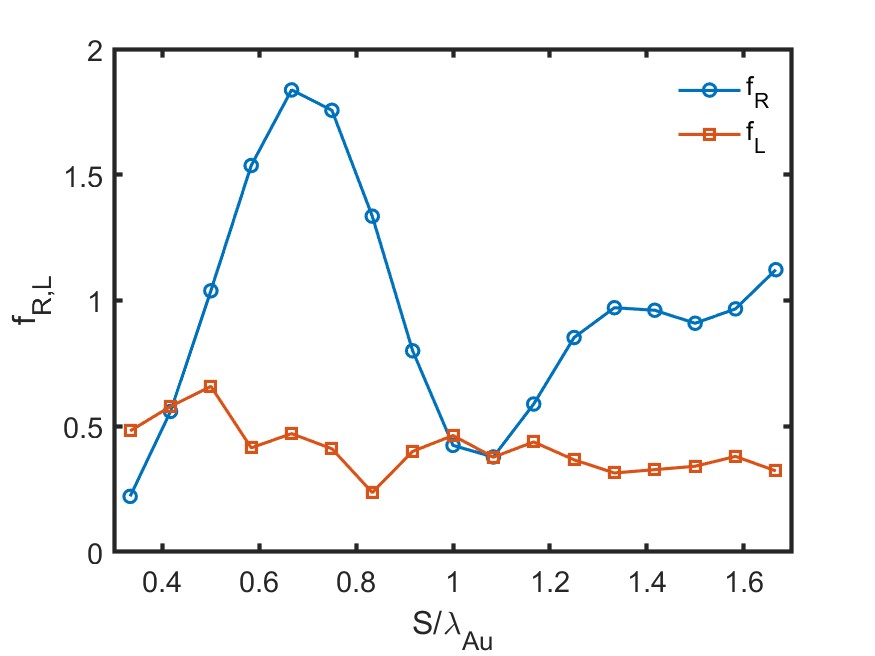}\\
    
    \end{tabular}
\caption{Directivity of silver (\textbf{a}) and gold (\textbf{b}) CR antennas with and without NP coupled to the dipole emitter.  Purcell ($f_R$)  and loss ($f_L$) factors  for silver {(\textbf{c}) and gold (\textbf{d}) CRs}
 with a single dipole emitter feeder.\label{fig::CRsolo}}
\end{figure}

The incorporation of the CR in the proposed hybrid NAs is targeted to shape the radiation produced by the dipole-NP feeder system.  It is important to evaluate the possible interaction effects between the CR and the feeder.  First, we calculated the properties of a $W=80$ nm thick CR when the emitter-NP was replaced by a single dipole emitter feeder.  \mbox{Figure \ref{fig::CRsolo}a,b} compare the the directivities as a function of $S/\lambda_M$ obtained with the CR  NA with and without the spherical NP coupled to the dipole emitter.  In the latter case, the position of the dipole emitter was centered at the origin.  The results demonstrate that the spatial shaping of the emission obtained by using the CR is largely independent of the dipole feeding system employed, since the differences between the directivities obtained with and without the NP were negligible.  Figure \ref{fig::CRsolo}c,d display the Purcell and loss factors for silver and gold nanoantennas, respectively, in the same cases as the previous figures.  The losses due to the CRs were much smaller that those obtained for a dipole coupled to a metal nanosphere and showed little dependence with respect to the position of the dipole emitter along the corner axis.  For the case of silver CRs, the losses were negligible at $\lambda=450$ nm.  As regards the Purcell factor, we observed the same type of pattern as that shown in \mbox{Figures \ref{fig::cornerplata} and \ref{fig::corneroro}} regarding the variation in $f_R$ with $S/\lambda_M$ due to the interaction of the dipole emitter with the CR plasmonic antenna, but the deviations from the unity were milder that those of the NP--emitter system and much smaller than those of the NP--emitter incorporated as a feeder system in the CR NA.  In summary, whereas the CR alone was responsible for managing the directive properties of the hybrid NA, the strong plasmonic interactions observed in the system, with large variations in $f_R$, were found to be mediated by the spherical NP at the \mbox{feeder arrangement}.

\begin{figure}[H] 
    \begin{tabular}{cc}
    {(a)}&{(b)}\\
    \includegraphics[width=6 cm]{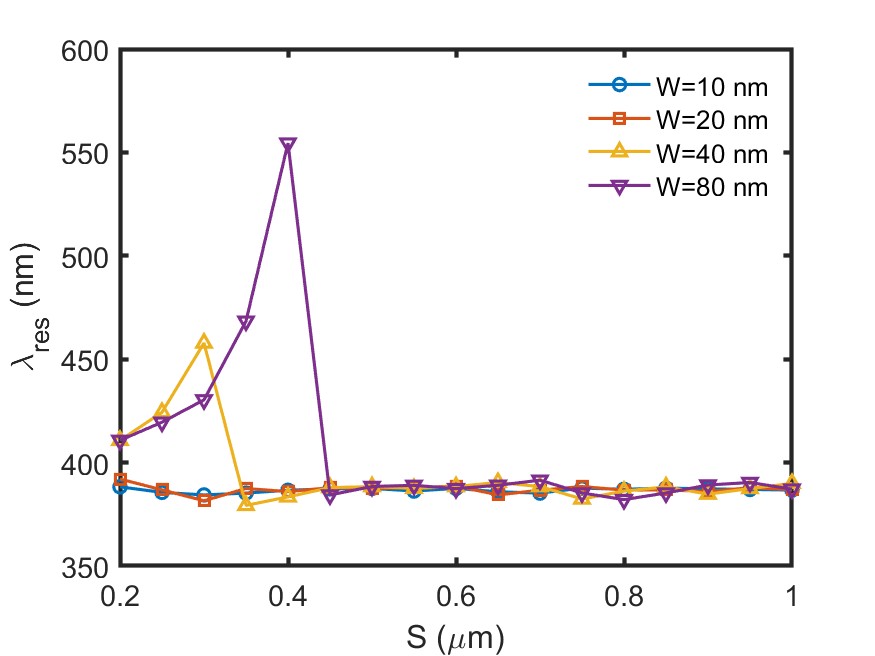}&
    \includegraphics[width=6 cm]{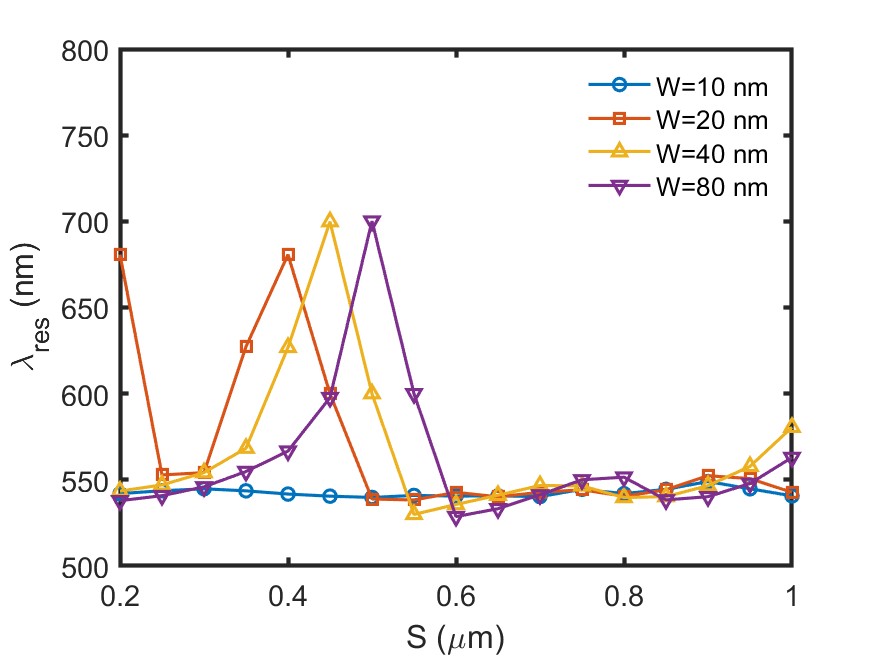}
    \end{tabular}
  \caption{Calculated resonance wavelengths of silver (\textbf{a}) and gold (\textbf{b}) NPs as a function of the absolute feeder position $S$ for four different reflector film thicknesses:  $W=10$ nm, $W=20$ nm, $W=40$ nm, and $W=80$ nm.\label{fig::lambdassin}}
\end{figure}

Figure \ref{fig::lambdassin} displays the plasmonic resonance wavelengths when the NPs were incorporated into the CR setup. These results illustrate the impact of NP--CR plasmonic interactions on the Purcell factor. Note that the distances between the feeder NP and the corner apex are unnormalized in these plots. The results obtained for spherical NPs, shown in Figure \ref{fig::resonancias}, exhibit minimal variation in the resonance wavelengths with respect to particle size and the distance between the NP and the emitter. In contrast, Figure \ref{fig::lambdassin} demonstrates that, for sufficiently large mirror thicknesses $W$, there were significant shifts in the resonance wavelengths as the position of the feeder emitter--NP was moved along the corner axis. As $S$ increased, there was an initial rise in $\lambda_{res}$, followed by a sudden drop to approximately the resonance wavelength of the isolated NPs. This drop occurred for values of $S$ near the minimum $f_R$ in \mbox{Figures \ref{fig::cornerplata} and \ref{fig::corneroro}}, highlighting the relationship between these results and the interaction properties of the CR with the NP in close proximity.

The modification of the emitter’s fluorescence rate by the NA system is directly related to the local field enhancement (LFE) at the emitter’s position due to the surrounding plasmonic environment \cite{novotny2006}. To evaluate the LFE factor, the NA setup must be in receiving mode. Reciprocity \cite{krauslibro,balanis} ensures that the antenna properties remain the same whether it is receiving or transmitting electromagnetic radiation. For these calculations, a linearly polarized plane wave, with the $E$ field aligned along the $Z$ axis, impinges on the NA along the $Y$ axis in the reverse direction. The local field enhancement factor can be evaluated as the ratio
\begin{equation}
  f_E=\dfrac{\left|\mathbf{E}\right|}{\left|\mathbf{E}_0\right|}
\end{equation}
between the field strength at the dipole source position $\left|\mathbf{E}\right|$ and the incident field strength $\left|\mathbf{E}_0\right|$, which is the field strength that would be found at that position in the absence of \mbox{the NA}.

\begin{figure}[H]   
    \begin{tabular}{cc}
    {(a)}&{(b)}\\
    \includegraphics[width=7 cm]{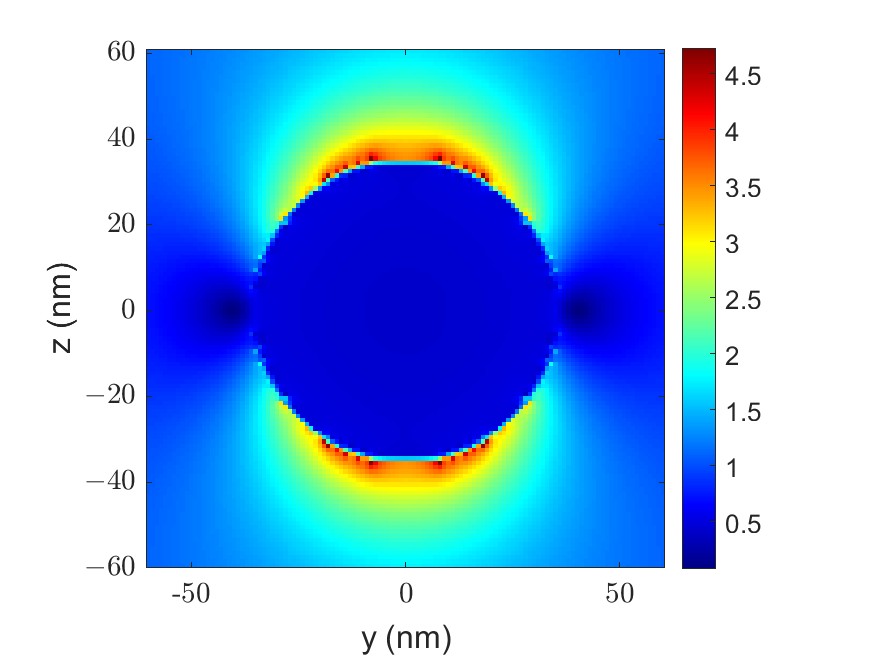}&
    \includegraphics[width=7 cm]{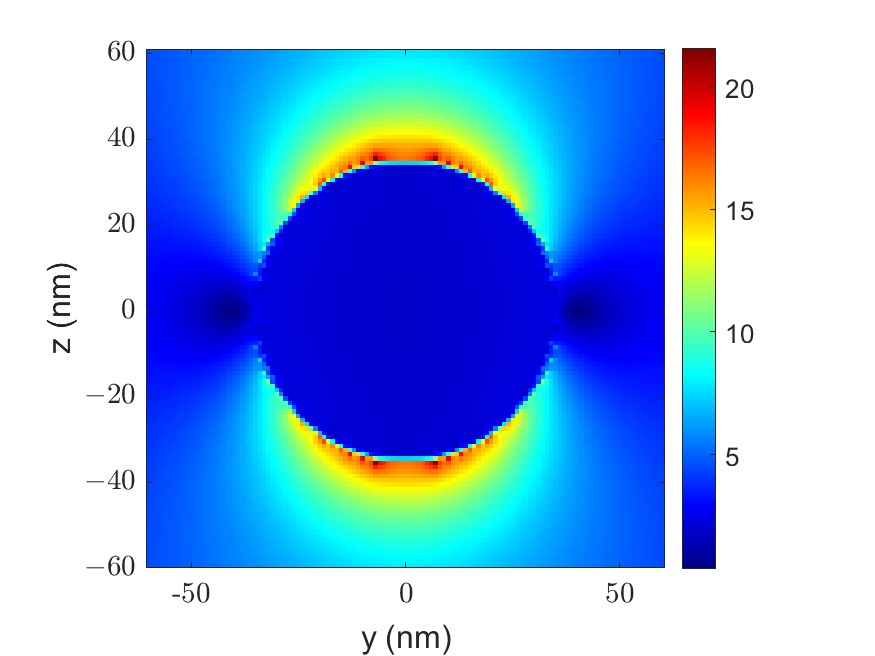}
    \end{tabular}
\caption{Field strength $|\mathbf{E}|$ across the $XZ$ plane in the environment of a $d=70$ nm gold NP at the origin illuminated by an unit magnitude $Z$-polarized E-field plane wave.  The plot (\textbf{a}) displays the results for the isolated NP and (\textbf{b}) when the NP was within the CR at $S=0.35$ $\upmu$m.}  \label{fig::LFE}
\end{figure}

Figure \ref{fig::LFE}a displays the calculated electric field strength $|\mathbf{E}|$ over the $XZ$ plane scattered from a gold nanosphere when illuminated by a unit amplitude E-field plane wave. For the transmitting mode of the NA, the dipole source was positioned at $x=0$, $z=50$ nm. Figure \ref{fig::LFE}b shows the corresponding electric field amplitude for the CR NA at the first directivity maximum. The results reveal a significant LFE for the isolated NP and a strong enhancement of this effect by the CR at the selected feeder position $S=0.35$ $\upmu$m in agreement with the values obtained for $f_R$. The calculated LFE factor for the previously optimized geometries yielded, for silver NAs, $f_E=13.8$ at $S=0.3$ $\upmu$m and $f_E=12.8$ at $S=0.55$ $\upmu$m. For gold NAs, the values were $f_E=9.9$ at $S=0.35$ $\upmu$m and $f_E=9.1$ at \mbox{$S=0.75$ $\upmu$m}. For the isolated NPs, the calculated LFE factors were $f_E=2.8$ for silver and $f_E=2.2$ for gold nanospheres.

\unskip

The properties of plasmonic NAs are largely influenced by interactions with nearby metal nanostructures. These can be leveraged in designing more efficient emitting structures, such as nanoparticle dimers \cite{dipole, chowdhury} or optimized NP arrays \cite{yagui1, yagui2}, and are also relevant in the development of substrates for surface-enhanced fluorescence spectroscopy \cite{sef} and surface-enhanced Raman scattering (SERS) applications \cite{sers}. We now examine the potential effects on the emission properties of the CR NA due to a nearby metal NP. To this end, we considered the presence of a $d=70$ nm NP at a distance $\rho=1.5$ $\upmu$m from the feeder NP at the origin, which was positioned at various angles $\alpha$ from the $Y$ axis. The position vector of the NP is given by $\mathbf{r}_n=(\rho\sin(\alpha_n), \rho\cos(\alpha_n), 0)$, where the values of $\alpha$ considered were $\alpha_n = n \cdot 30^\circ$ for $n = 0 \dots 6$.

\begin{figure}[H]   
    \begin{tabular}{cc}
    {(a)}&{(b)}\\
    \includegraphics[width=6 cm]{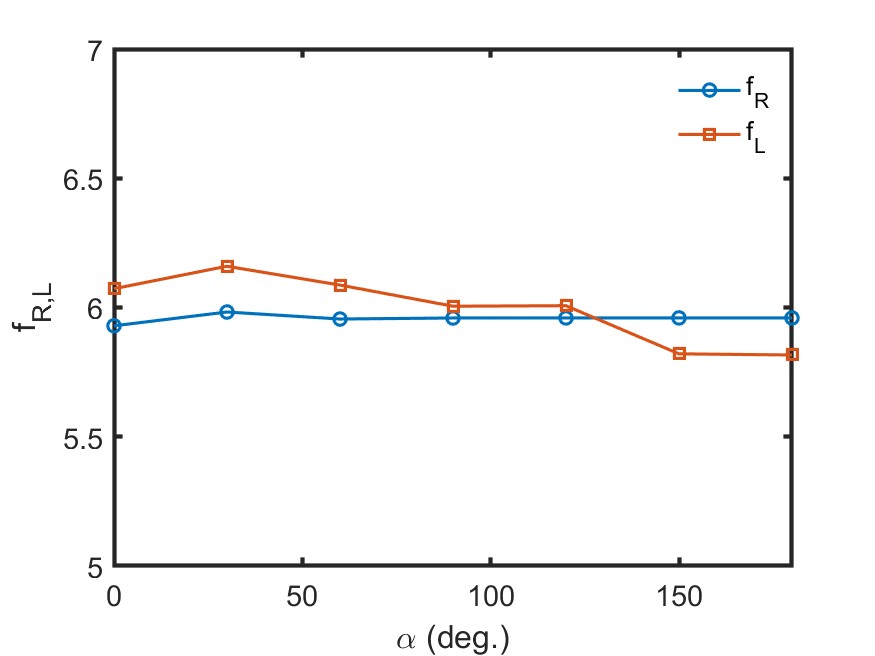}&
    \includegraphics[width=6 cm]{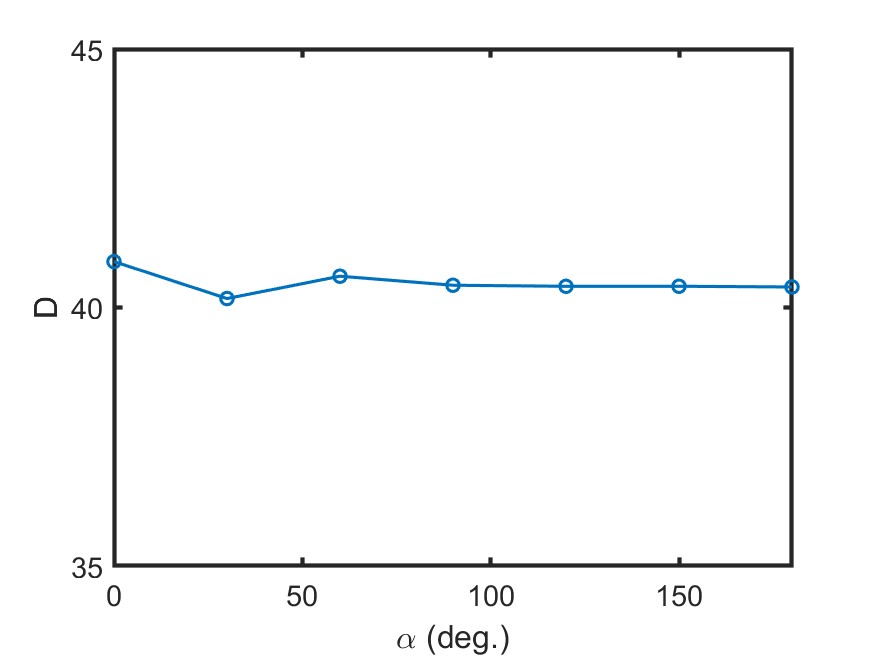}
    \end{tabular}
\caption{(\textbf{a}) Purcell, $f_R$, and loss, $f_L$, factors; (\textbf{b}) directivity of a $W=80$ nm, $L=1.5$ $\upmu$m CR silver NA with the feeder at $S=0.55$ $\upmu$m in the presence of an interfering $d=70$ nm silver NP at position $(\rho\sin(\alpha),\rho\cos(\alpha),0)$ for different values of $\alpha$. \label{fig::shield}}
\end{figure}

Figure \ref{fig::shield} shows the results calculated for a silver CR NA with $W=80$ nm in thickness and a feeder position of $S=0.55$ $\upmu$m in the presence of an interfering NP. The results indicate a very small deviation in the values of $f_R$ and $f_L$ compared to those determined without the interfering NP, as shown in Figure \ref{fig::cornerplata}d, regardless of the position angle $\alpha$. The directivity values in Figure \ref{fig::shield}b also exhibit minimal deviations from those obtained without the interfering NP, shown in Figure \ref{fig::directividades}a, especially when the interfering NP was positioned behind the CR. These calculations demonstrate a certain degree of robustness in the CR NA's response to plasmonic NP contamination in the NA surroundings, at least outside the region defined by a circle centered at the feeder NP with a radius equal to the mirror length $L$ of the CR.

\begin{figure}[H]   
    \begin{tabular}{cc}
    {(a)}&{(b)}\\
    \includegraphics[width=6 cm]{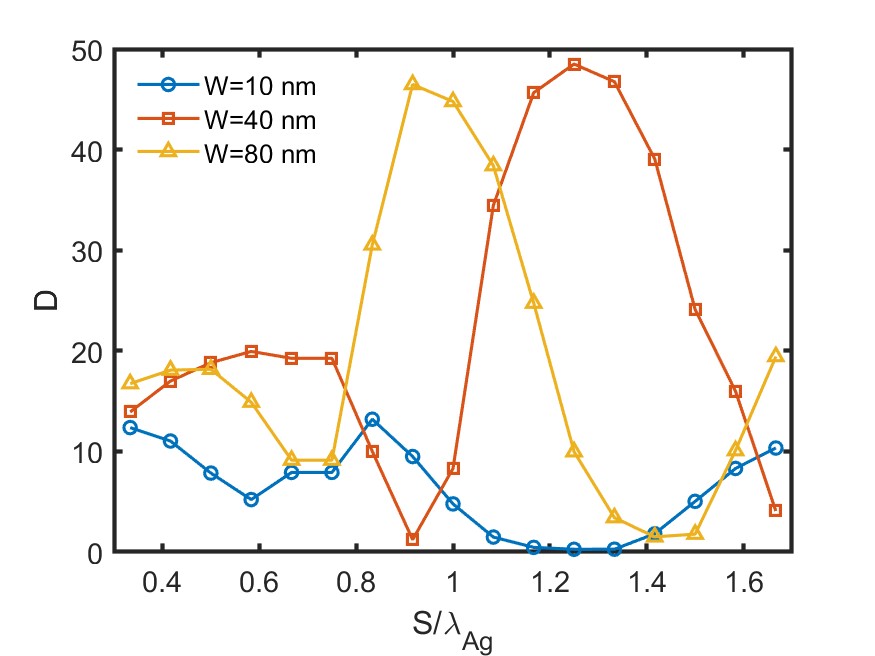}&
    \includegraphics[width=6 cm]{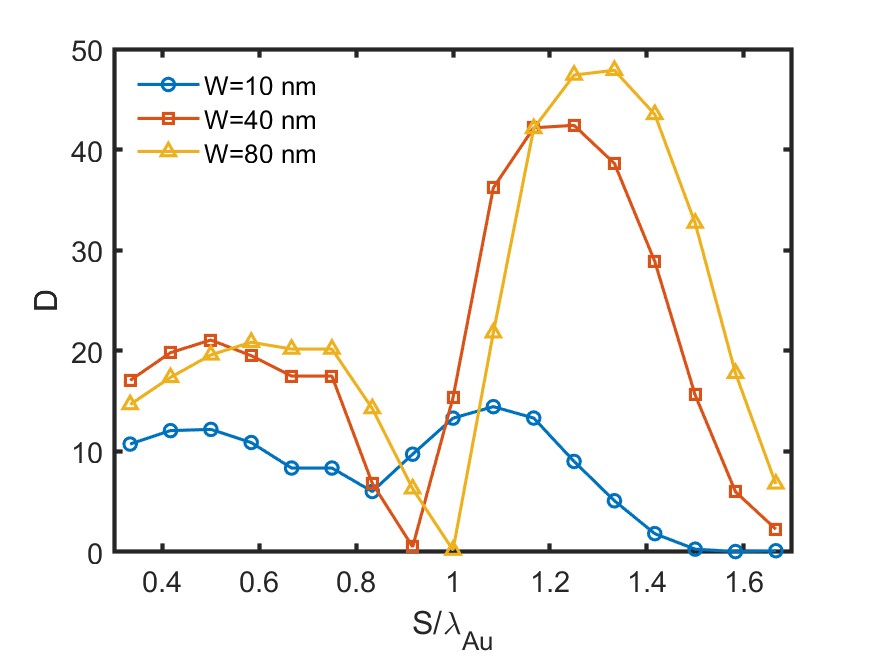}
    \end{tabular}
\caption{Values obtained of the directivity as a function of the feeder position $S$ normalized to the corresponding target wavelength for three different reflector film thicknesses when the length of the CR planes was increased to $L=3$ $\upmu$m. (\textbf{a}) Results for silver CR.  (\textbf{b}) Results for gold CR.  \label{fig::directividades3um}}
\end{figure}

Figure \ref{fig::directividades3um} shows the directivities calculated for the silver and gold CR NAs when the corner length was increased to $L=3$ $\upmu$m for three values of $W$ ($10$ nm, $40$ nm, and $80$ nm).  For this $L$, the peak directivities were essentially the same for the silver and gold CRs, even though the $L/\lambda_{M}$ value was larger for silver.  This indicates a saturating effect regarding the increase in $D$ with $L$.

In the case of gold, the relative positions of the maxima and minima as a function of $S/\lambda_M$ closely matched those found for $L=1.5$ $\upmu$m, with a null occurring at $S=\lambda_M$ for $W=80$ nm.  In contrast, for silver, significant shifts in the directivity maxima and minima as a function of $S/\lambda_M$ were observed , and these shifts were highly dependent on the value \mbox{of $W$.}

\begin{figure}[H]   
    \begin{tabular}{cc}
    {(a)}&{(b)}\\
    \includegraphics[width=6 cm]{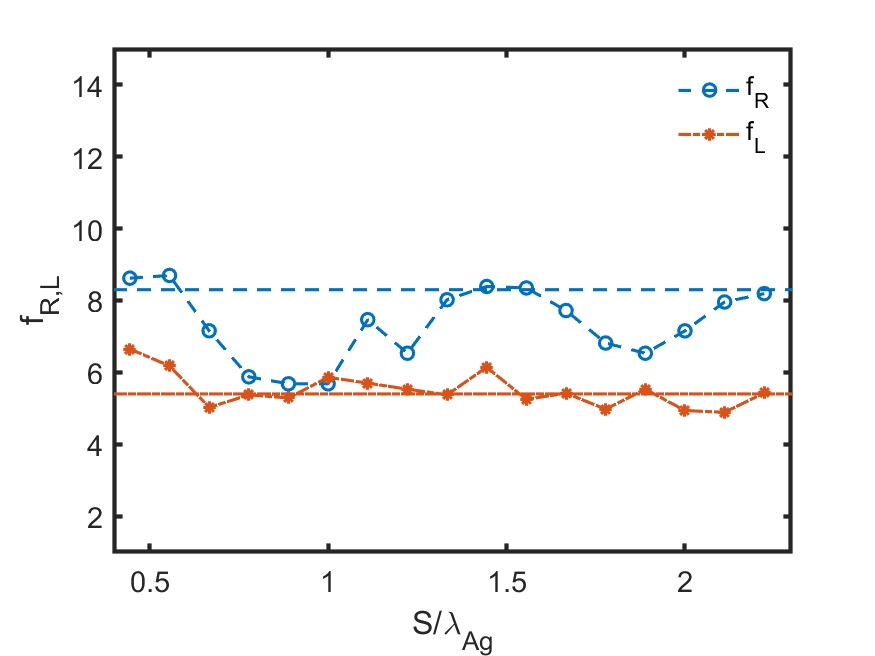}&
    \includegraphics[width=6 cm]{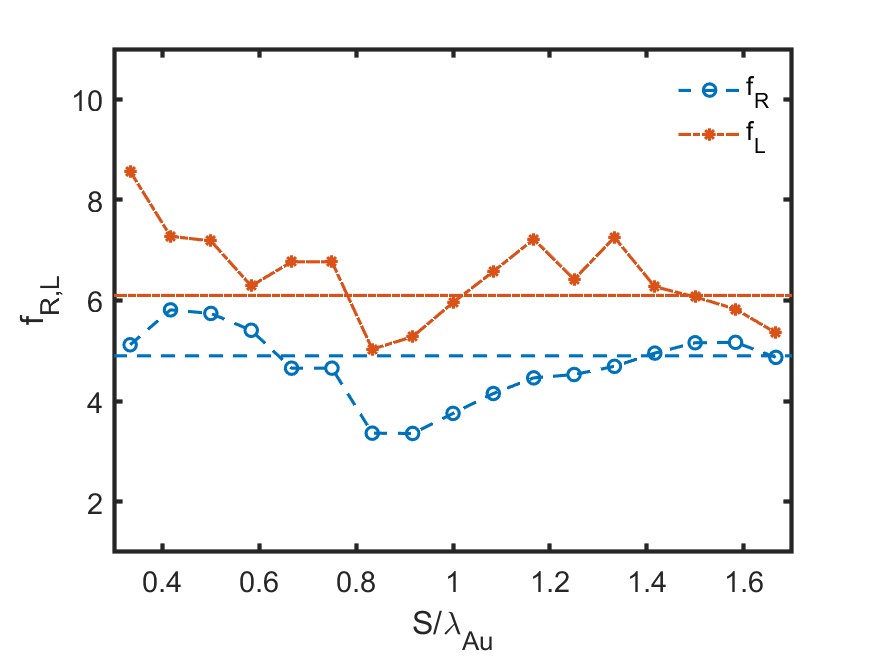}\\
     {(c)}&{(d)}\\
    \includegraphics[width=6 cm]{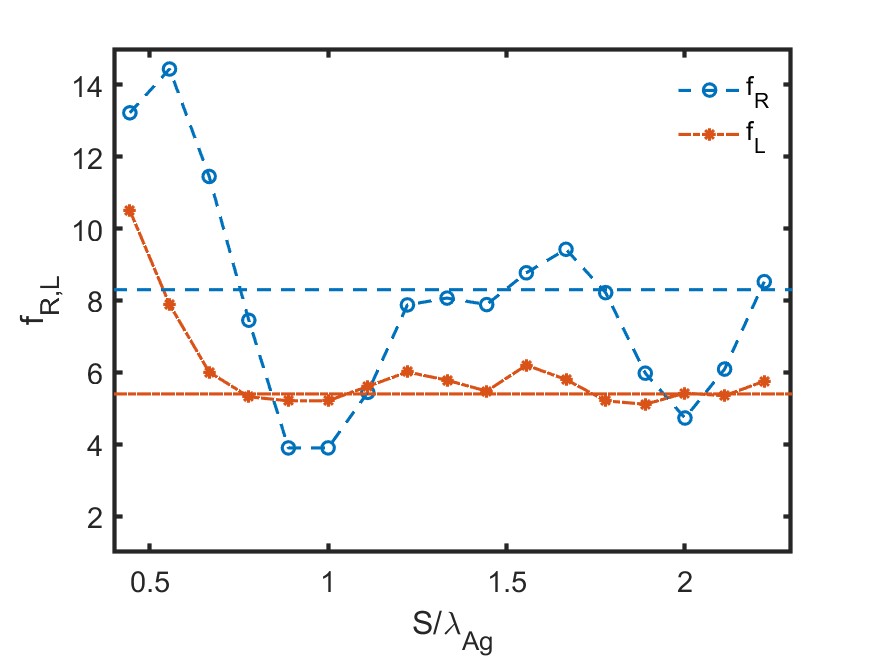}&
    \includegraphics[width=6 cm]{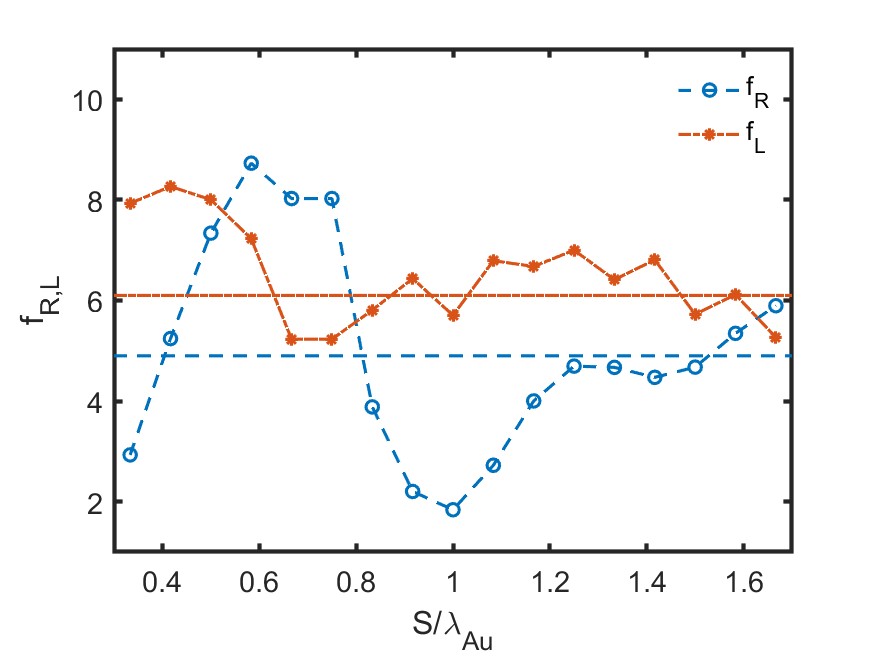}\\
    {(e)}&{(f)}\\
    \includegraphics[width=6 cm]{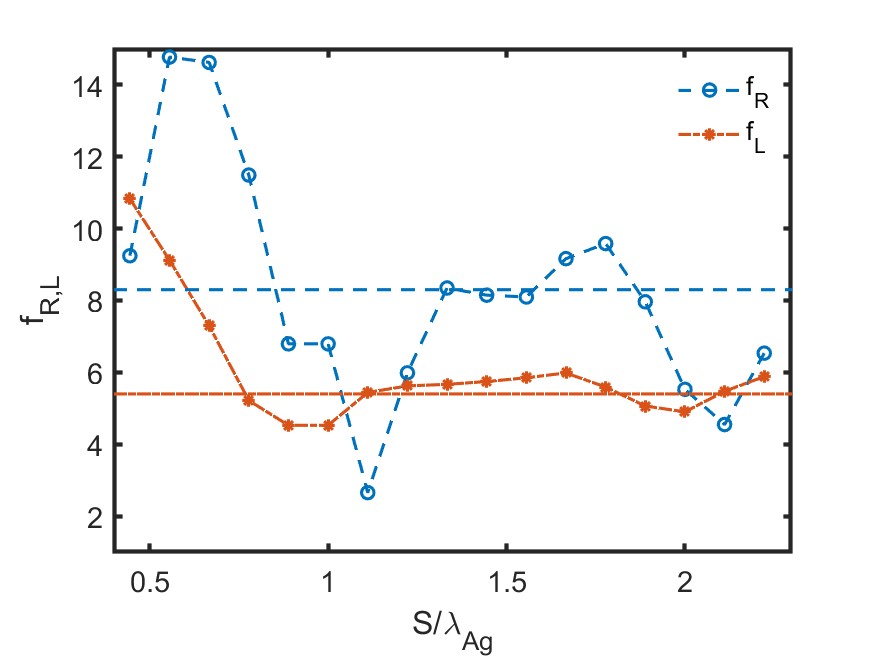}&
    \includegraphics[width=6 cm]{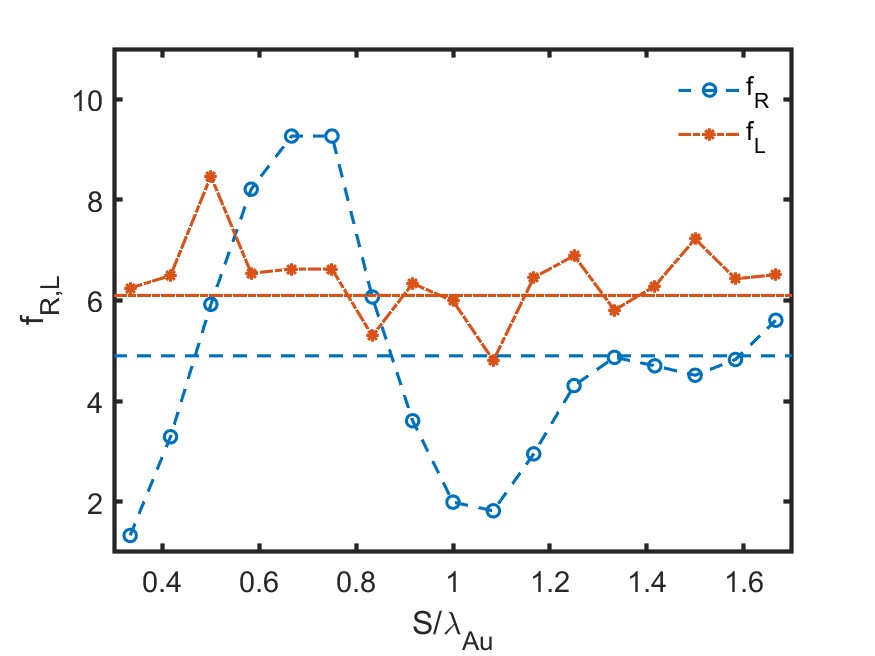}
    \end{tabular}
  \caption{Radiative transition rate enhancement $f_R$ and normalized loss $f_L$ factors of a gold CR with $L=3$ $\upmu$m as a function of the feeder position $S/\lambda_M$ for three different reflector film thicknesses: \mbox{(\textbf{a}) silver} CR $W=10$ nm, (\textbf{b}) gold CR $W=10$ nm, (\textbf{c}) silver CR $W=40$ nm, (\textbf{d}) gold CR $W=40$ nm, (\textbf{e}) 
 silver CR $W=80$ nm, (\textbf{f}) gold CR $W=80$ nm.\label{fig::corner3um}}
\end{figure}

The normalized radiative emission and loss rates for $L=3$ $\upmu$m CRs are displayed in Figure \ref{fig::corner3um} for the three values of $W$ considered.  The results closely resemble those obtained for $L=1.5$ $\upmu$m, indicating that the impact of the CR on the loss and emission properties was due to the interactions occurring in a localized region near the apex of the CR.

\begin{figure}[H]  
    \begin{tabular}{cc}
    {(a)}&{(b)}\\
    \includegraphics[width=6 cm]{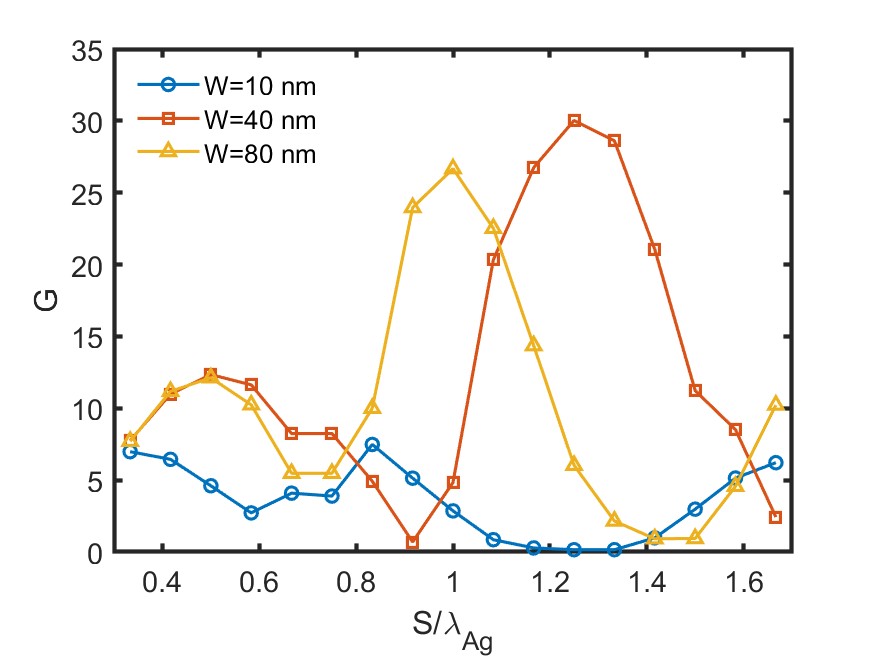}&
    \includegraphics[width=6 cm]{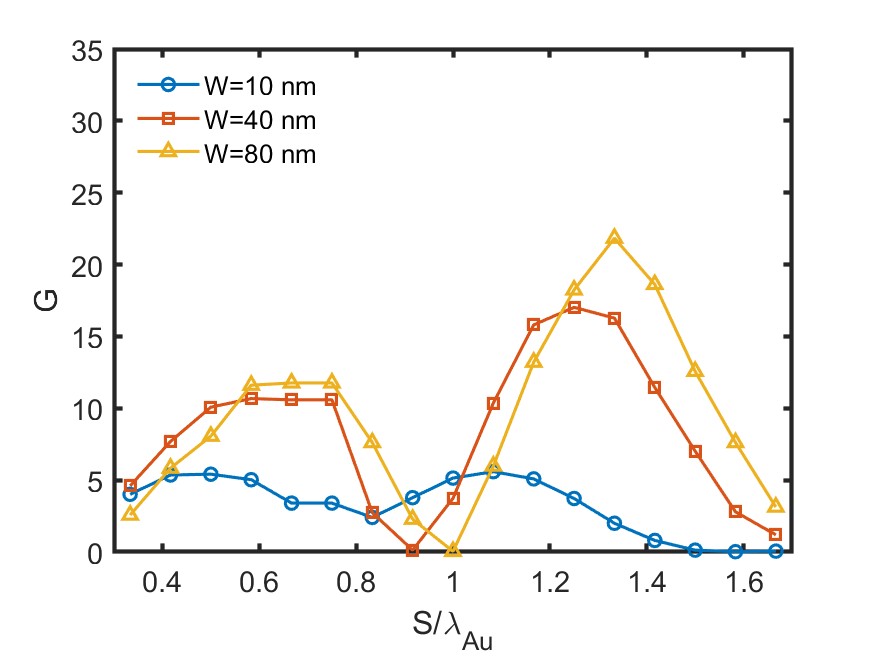}
    \end{tabular}
\caption{Nanoantenna gain as a function of the feeder position $S$ normalized to the corresponding target wavelength for three different reflector film thicknesses when the length of the CR planes was increased to $L=3$ $\upmu$m. (\textbf{a}) Results for silver CR.  (\textbf{b}) Results for gold CR.  \label{fig::G3um}}
\end{figure}

Figure \ref{fig::G3um} shows the gain, $G$, of the silver and gold CR NAs as a function of $S/\lambda_M$.  The gain provides a comprehensive measure of on-axis photon emission, incorporating all three key parameters: $f_R$, $f_L$, and $D$.  For the gold CR NAs, the maximum gain occurred at a metal thickness of $W=80$ nm.  The first gain peak, with a value of $G=11.8$, was found at $S/\lambda_{Au}=0.67$ and $S/\lambda_{Au}=0.75$.  The second peak, $G=21.8$, was located at $S/\lambda_M=1.33$.

In the case of the silver CRs,  the aforementioned shift of the directivity maxima with respect to $S/\lambda_{Ag}$ resulted in the highest directivity for $W=40$ nm rather than for $W=80$ nm.  This shift aligns the directivity peak with a higher $f_R$ value at $W=40$ nm, leading to the maximum gain being achieved at this thickness.  The first gain peak for silver was $G=12.3$ at $S/\lambda_{Ag}=0.5$, and the second was $G=30.0$ at $S/\lambda_{Ag}=1.25$.

  The directivity values obtained for the CR NA were, in all cases, well below the bounds set by the Kildal limit for the CR sizes considered \cite{kildal}. Nevertheless, the proposed scheme offers a significant improvement over a single NP emitter while maintaining high implementation simplicity, which is consistent with the advantages of CR antennas used in radio frequency applications. Although superdirective dielectric NAs would allow for a more compact setup, the proposed scheme enables the full exploitation of plasmonic nanostructures' potential \cite{khurgin}.

\section{Conclusions}

The analysis of dipole--NP interactions, combined with CR integration, reveals important insights into the enhancement of radiative properties for silver and gold NPs. Knowing the fundamental behavior of NP interactions is crucial for understanding the modifications introduced by the incorporation of the CR in the system.  Initially, isolated dipole--NP calculations without the CR showed trends consistent with previous theoretical and experimental studies, particularly regarding the variation in resonance wavelength, the radiative emission rate enhancement factor, and loss factor with silver and gold particle size and dipole emitter separation.  This preliminary study established a set of parameters for the subsequent analysis of the CR. The results also suggest that silver NPs are more effective than gold NPs in enhancing radiative emission while maintaining lower loss factors.

Integrating CRs into silver and gold NA systems leads to substantial improvements in directivity.  The variation in directivity with respect to the normalized distance $S/\lambda_M$ reveals that thicker metal films generally lead to more regular oscillatory patterns in directivity. At the first directivity maxima, significant gains in the radiative enhancement factor, $f_R$, were also achieved with minimal increases in the loss factor, $f_L$, leading to high gain figures of merit, $G$. At the second directivity maximum, the CR mainly enhanced directivity, with the highest values of $D$ corresponding to the global maxima of $G$.

Extending the corner length to $L=3$ $\upmu$m demonstrated a saturation effect in the enhancement of directivity for both materials, with similar peak values achieved for the gold and silver CRs despite differences in their $L/\lambda_M$ values.  Meanwhile, the patterns for the variations in $f_R$ and $f_L$ closely resembled those found at $L=1.5$ $\upmu$m.  These findings highlight the localized nature of CR-induced enhancements near the apex.  However, silver CRs exhibited larger shifts in their positions of the directivity maxima and minima with changes in $W$.  This resulted in the maximum gain for silver CR NAs at a metal thickness of $W=40$ nm, rather than \mbox{$W=80$ nm}, due to the alignment of the directivity maximum with a higher value of $f_R$.

Overall, the results indicate that silver CR NAs outperform gold CR NAs in terms of radiative enhancement and gain, offering more efficient light emission and directivity control for nanoscale optical applications, although different spectral windows are targeted by the choice of NA metal.

The reflector angle selected for the analysis aligns with a typical value used in RF applications, corresponding to optimization for operation at the more favorable second gain maximum. The analysis reveals a strong metal--feeder interaction near the corner apex, suggesting the potential benefits of further exploring corner geometries to optimize the first gain maximum, which will be addressed in future works.

\section*{Acknowledgement}

This work was funded by the Spanish Ministerio de Ciencia e Innovaci\'on (MCIN) under grant PID2020-119418GB-I00, by the European Union NextGeneration EU under grant PRTRC17.I1, and by the Consejería de Educación, Junta deCastilla y León, through the QCAYLE project and grant VA184P24.

\end{document}